\documentclass[final,5p,times,twocolumn]{elsarticle}

\usepackage{amsmath,epsfig,amstext,amsfonts,amssymb}
\usepackage{ae,aecompl} 
\usepackage{amscd,amsmath,amssymb,amsfonts,latexsym,amsthm,mathrsfs,bbm,ifthen,enumerate}
\usepackage{amsmath,amstext,amsfonts,amssymb,color}
\usepackage[mathscr]{euscript}
\usepackage{graphicx,subfigure}
\usepackage{multirow}
\usepackage{slashbox,tabularx}
\usepackage{relsize}
\usepackage{subfigure}
\usepackage{slashbox}
\usepackage{pifont}

\usepackage{subfigure}

\newtheorem{corollary}{Corollary}

\newtheorem{lemma}{Lemma}

\newtheorem{remarque}{Remark}

\newtheorem{proposition}{Proposition}

\newtheorem{theorem}{Theorem}

\newcommand{\bs}{\boldsymbol}

\def\EE{ {\mathbb{E}}}

\def\LLL1{ L^1({\mathbb{R}})  }
\def\LL2{ L^2({\mathbb{R}})  }
\def\Ll1{ l^1({\mathbb{Z}})  }

\journal{}

\begin{document}

\begin{frontmatter}



\title{Blind source separation of convolutive mixtures of non circular linearly modulated
signals with unknown baud rates.}


\author[ad-ef]{E. Florian}
\author[ad-ac]{A. Chevreuil }
\author[ad-ef]{P. Loubaton}
\address[ad-ef]{Université Paris-Est, LIGM Laboratoire d'Informatique Gaspard Monge, UMR CNRS 8049 \\
5 Bd. Descartes, Cité Descartes, 77454 Marne la Vallée  Cedex 2}
\address[ad-ac]{ESIEE Paris, 2Bd. BlaisePascal, Cité Descartes, BP99, 93162 Noisy-le-Grandcedex, France}

\begin{abstract}
  This paper addresses the problem of blind separation of convolutive mixtures of BPSK and circular linearly modulated signals with unknown (and possibly different) baud rates and carrier frequencies. In previous works, we established that the Constant Modulus Algorithm (CMA) is able to extract a source from a convolutive mixture of circular linearly modulated signals. We extend the analysis of the extraction capabilities of the CMA when the mixing also contains BPSK signals. We prove that if the various source signals do not share any non zero cyclic frequency nor any non conjugate cyclic frequencies, the local minima of the constant modulus cost function are separating filters. Unfortunately, the minimization of the Godard cost function generally fails when considering BPSK signals that have the same rates and the same carrier frequencies. This failure is due to the existence of non-separating local minima of the Godard cost function. In order to achieve the separation, we propose a simple modification of the Godard cost function which only requires knowledge of the BPSK sources frequency offsets at the receiver side.
We provide various simulations of realistic digital communications scenarios that support our theoretical statements.    
\end{abstract}


\end{frontmatter}

%
%
%
%
%
%
%
%
%
%
%
%
%
\section{Introduction}

\label{sec:intro}
The blind source separation of convolutive mixtures of linearly modulated signals has mainly been studied in the case where the signals share the same known baud rate, and when the sampling frequency of the multivariate received signal coincides with this baud-rate. In this context, to be referred to in the sequel as the {\em stationary case}, the discrete-time received signal 
coincides with the output of an unknown MIMO filter driven by the sequences of symbols sent by the various transmitters. 
In most cases, these sequences are independent and identically distributed, and several methods have been proposed in order 
to extract each of them from the observation (see e.g. \cite{comon-1996}, \cite{inouye-1998}, \cite{inouye-liu-2002}, \cite{tugnait-ieeesp-97}, \cite{tugnait-ieeesp-1997-bis}) . The source separation problems that are encountered in the context of passive listening 
are however more complicated because the transmitters are usually completely unknown to the receiver, and have no reason to 
transmit linearly modulated signals sharing the same baud-rates. It is therefore quite relevant to address the problem of blind separation 
of linearly modulated signals with unknown, and possibly different, baud rates. In this context, the received signal is sampled at 
any frequency satisfying the Shannon sampling theorem, so that the corresponding discrete-time signal is cyclostationary with unknown 
cyclic frequencies. If the cyclic frequencies were known at the receiver side, it would be easy to generalize 
the usual blind source separation approaches based on the optimization of contrast functions depending on higher order cumulants. However, when the 
cyclic frequencies are unknown, it is impossible to consistently estimate the cumulants, a conceptual problem first remarked by 
Ferreol and Chevalier (\cite{art:che-1}) in the context of blind separation of instantaneous mixtures. An obvious approach would consist in estimating
the unknown cyclic frequencies. However, this is a difficult task if the excess bandwidths of the 
transmitted signals are low and if the duration of observation is not large enough. 


In contrast with the cumulants, the constant modulus cost function can be consistently estimated in the cyclostationary context.
In \cite{jal-che-lou-06-soumis}, we considered only source signals that transmit second-order circular symbol sequences, and we have shown that in this case, to be referred to as the \emph{circular case}, the minimization of the Godard cost function allows to extract the sources using a deflation approach if their baud-rates are different one from another. If certain baud rates coincide, sufficient conditions for the separation have been established in \cite{jal-che-lou-06-soumis}. Although we have not been able to prove that separation is achieved in the most general case, all the simulations we have performed strongly suggest that the minimization of the Godard cost function is successful in the circular case.
The purpose of this paper is to address this issue when in the non circular source signals, which will be referred to as the {\emph{non circular} case}, and to show how the separation method based on the minimization of the CMA contrast function coupled with a deflation approach can be adapted to this context. As in \cite{jal-che-lou-06-soumis} we only focus in this paper on the separation of the first source. 

In order to simplify the presentation of our results, we only consider the case where the non circular signals are BPSK signals. 
We begin by defining in section \ref{sec:signal-model} the context of our study and giving a brief description of the considered signals and criteria. In section \ref{sec:differentes} we prove that the Godard cost function is still successful if the sources do not share the same baud rates and the same carrier frequencies. We also prove, in section \ref{sec:egales}, that contrary to the circular case, the minimization of the Godard cost function fails to separate 2 BPSK signals sharing the same baud rate and the same carrier frequency. We show that this is due to the existence of non separating local minima of the Godard cost function, toward which the minimization algorithms seem to converge quite often. 
We also show that it is possible to modify the Godard cost function in order to achieve source separation of $K$ non circular BPSK modulated signals sharing the same known (or well estimated) carrier frequency. Section \ref{sec:Rem} briefly generalizes this result to more general mixtures. The new modified CMA algorithm needs the estimation of the carrier frequencies offsets of the non circular source signals, or equivalently the estimation of the "significant" non conjugate cyclic frequencies of the received signal. Fortunately, this is a much easier task than the estimation of baud rates, because the non conjugate cyclic correlation coefficients of the received signal at twice the frequency offsets are not affected by possible low excess bandwidths of the source signals (see \cite{ciblat-ieeesp-2002}). Numerical results are finally  presented in section \ref{sec:sim}. 


{\em Notations:} If $(u_n)_{n \in \mathbb{Z}}$ is a discrete-time sequence, we denote by $< u_n>$ the time average operator defined as
$$
< u_n > = \lim_{N \rightarrow + \infty} \frac{1}{2N + 1} \sum_{n=-N}^{N} u_n
$$
If $x$ is a complex valued random variable, we denote by $c_4(x)$ its fourth order cumulant defined by $cum\{x,x^{*},x,x^{*}\}$.
If $(x(n))_{n \in \mathbb{Z}}$ is a discrete-time cyclostationary sequence, we define, when it makes sense, the cyclo-correlation at cyclic-frequency $\alpha$ and time lag $m$:
$$\forall \alpha\in \left(-\frac{1}{2},\frac{1}{2}\right],\, \forall m\in\mathbb{Z},\, R^{(\alpha)}_x(m)= <\mathbb{E}(x(n+m)x(n)^{*}e^{-2i\pi n\alpha})>$$
and the non conjugate cyclo-correlation at cyclic-frequency $\alpha_c$ and time lag $m$:
$$\forall \alpha_c\in \left(-\frac{1}{2},\frac{1}{2}\right],\, \forall m\in\mathbb{Z},\, R^{(\alpha_c)}_{c,x}(m)= <\mathbb{E}(x(n+m)x(n)e^{-2i\pi n\alpha})>$$

For a wide-sense cyclostationary continuous-time random process $(x_a(t))_{t\in\mathbb{R}}$ we denote by  $R^{(\alpha_a)}_{a,x}(\tau)$ and by $R^{(\alpha_{a,c})}_{a,c,x}(\tau)$ the cyclic correlation coefficient and respectively non conjugate cyclic correlation coefficient at cyclic-frequency $\alpha_a$(respectively non conjugate cyclic frequency $\alpha_{a,c}$) and time lag $\tau$.

For an interval $\mathcal{B}$, we denote by $\mathcal{F}(\mathcal{B})$ the set of all functions $f_a(t)\in \mathbb{L}^2(\mathbb{R})$ such that
$$ f_a(t)=\int_{\mathcal{B}}{s^{2i\pi\nu t}\hat{f}_a(\nu)}d\nu$$ 
In other words, a square integrable function $f_a$ is an element of $\mathcal{F}(\mathcal{B})$ if and only if its Fourier transform $\hat{f}_a(\nu)$
is zero outside $\mathcal{B}$. 

\section{Problem statement} 
\label{sec:signal-model}
\subsection{Assumptions}
We assume that $K$ unknown transmitters send linearly modulated signals sharing the same frequency bandwidth. The receiver is equipped with a sensor of $N$--arrays, and the corresponding $N$--dimensional received signal 
is sampled at rate $T_e$ supposed to satisfy the Shannon sampling theorem. For any $k$, $k= 1, \ldots, K$, the signal transmitted by source $k$ 
is obtained by linearly modulating a unit variance zero mean i.i.d. sequence of symbols $\{a_{k,n}\}_{n\in\mathbb{Z}}$ with a shaping filter $g_{a,k}$
$$ s_{a,k}(t)=\sum_{n\in\mathbb{Z}}{a_{k,n} g_{a,k}(t-nT_k)} $$
We denote by $T_k$ the symbol period of the source number $k$ and we consider a shaping filter of limited bandwidth $[-\frac{1+\gamma_k}{2T_k}, \frac{1+\gamma_k}{2T_k}]$, where $\gamma_k$ is the excess bandwidth factor, belonging to $[0, 1)$.
The bandwidth of the complex envelope of transmitted signal $k$ is then $[-\frac{1+\gamma_k}{2T_k}, \frac{1+\gamma_k}{2T_k}]$.

In order to simplify the presentation of the results we make the following assumption:
\begin{itemize}
\item the symbol sequence $\{a_{k,n}\}_{n\in\mathbb{Z}}$ is either second order circular or corresponding to a BPSK constellation (i.e. equal to $\pm 1$) for each $k$. 
\end{itemize}

The propagation channels between each transmitter and the receiver are assumed to be frequency selective. Moreover, the carrier frequencies of the 
various transmitted signals of course do not coincide with the center frequency of the receive filter of the receiver. Hence, 
the contribution of each transmitted signal at the receiver side is corrupted by a frequency offset. The frequency offset associated to 
source $k$ is denoted by $\Delta f_k$. 

We denote by ${\bf y}_{a,k}(t)$ the $N$ dimensional continuous-time signal representing the contribution of the transmitted signal  $k$ to the received signal ${\bf y}_{a}(t)$ which is to say, the signal that would be received if only transmitter $k$ were active. We can then write ${\bf y}_{a,k}(t)$ as 
\begin{equation}
\label{eq:expre-y_{a,k}}
{\bf y}_{a,k}(t) = e^{2 i \pi \Delta f_k t} \left( {\bf h}_{a,k} * s_{a,k} \right)(t)
\end{equation}
where $*$ represents the convolution operator and where ${\bf h}_{a,k}$ is the $N$ dimensional channel impulse response  between source $k$  
and the multiple-sensors receiver. The presence of the frequency offset shifts the bandwidth of the ${\bf y}_{a,k}(t)$ signal with a factor equal to $\Delta f_k$, thus making it coincide with the interval $[-\frac{1+\gamma_k}{2 T_k} + \Delta f_k, \frac{1+\gamma_k}{2 T_k} + \Delta f_k]$. 

The continuous-time received signal (in the absence of noise) ${\bf y}_a(t) = \sum_{k=1}^{K} {\bf y}_{a,k}(t)$ is sampled at rate $T_e$ which is supposed to verify 
\begin{equation}
\label{eq:Shannon}
\frac{1}{2T_e} > \max_k \left( \frac{1+\gamma_k}{2} + |\Delta f_k| \right)
\end{equation}
Under these assumptions, the $N$-dimensional discrete-time received signal ${\bf y}(n)$ can be written as 
\begin{equation}
\label{eq:modeley}
{\bf y}(n) = \sum_{k=1}^{K} e^{2 i \pi n \delta f_k} \left( \sum_l {\bf h}_{k,l} s_k(n-l) \right)  = \sum_{k=1}^{K} e^{2 i \pi n \delta f_k} [{\bf h}_k(z)]s_k(n)  
\end{equation}
where for each $k$,  $s_k(n)$ represents the sampled version of transmitted signal $k$, and where ${\bf h}_k(z) = \sum_{l \in \mathbb{Z}} {\bf k}_{k,l} z^{-l}$ is the transfer function of the $1$-input / $N$ outputs discrete time equivalent channel between transmitter $k$ and the receiver. Finally, $\delta f_k$ is defined as $\delta f_k = \Delta f_k T_e$. 
\subsection{Expansion of the Godard cost function}

Due to the previously described context, each of the transmitted signals is cyclostationary and thus has a set of cyclic frequencies which are easily identified from the second order statistics of each signal. For all $k$, and for all $\tau \in \mathbb{R}$, the cyclic correlation function $t \rightarrow \mathbb{E}(s_{a,k}(t+\tau) s_{a,k}(t)^{*})$ and, for a BPSK signal, the non conjugate cyclic correlation function $t \rightarrow \mathbb{E}(s_{a,k}(t+\tau) s_{a,k}(t))$ are periodic of period $T_k$. Because of the limited bandwidth of $s_{a,k}$, the expansion in Fourier series of these two functions only involves frequencies  $0, \frac{1}{T_k}$ and $ -\frac{1}{T_k}$ of $s_{a,k}$.
\begin{align*}
&\mathbb{E}(s_{a,k}(t+\tau) s_{a,k}(t)^{*}) = R_{s_{a,k}}^{(0)}(\tau) +  R_{s_{a,k}}^{(\frac{1}{T_k})}(\tau) e^{2 i \pi \frac{t}{T_k}} +  R_{s_{a,k}}^{(-\frac{1}{T_k})}(\tau)e^{-2 i \pi \frac{t}{T_k}}\\
&\mathbb{E}(s_{a,k}(t+\tau) s_{a,k}(t)) = R_{c,s_{a,k}}^{(0)}(\tau) +  R_{c,s_{a,k}}^{(\frac{1}{T_k})}(\tau) e^{2 i \pi \frac{t}{T_k}} +  R_{c,s_{a,k}}^{(-\frac{1}{T_k})}(\tau)e^{-2 i \pi \frac{t}{T_k}}\\
\end{align*}
Note that when the excess bandwidth $\gamma_k$ is small, the cyclic correlation coefficients  at non-zero frequencies are clearly inferior to those corresponding to the zero cyclic frequency. 


We denote  $\alpha_k = \frac{T_e}{T_k}$ for $k=1, \ldots, K$. Then, it is clear that the non zero cyclic frequencies of the discrete time signal are $\pm \alpha_k$; if moreover $s_k$ is a BPSK signal, its non conjugate cyclic frequencies are $0, \pm \alpha_k$. From now on, we denote by $I$ and $I_c$ the set of all cyclic and non conjugate cyclic frequencies of ${\bf y}(n)$. We obtain immediately that
\begin{itemize}
\item $I = \{0, (\pm \alpha_k)_{k=1, \ldots, K} \}$ 
\item $I_c = \{ (2 \delta f_k, 2 \delta f_k \pm \alpha_k)_{k=1, \ldots, K} , s_k \; \mathrm{BPSK} \}$ 
\end{itemize}
In the following, we also denote $I^{*}$ the set of non zero cyclic frequencies of ${\bf y}(n)$.

In order to extract one of the source signals, $({\bf y}(n))_{n \in \mathbb{Z}}$ is filtered by a $N$--inputs / 1--output filter ${\bf g}(z)$ 
to produce the 1--dimensional signal $r(n) = [{\bf g}(z)]{\bf y}(n)$. It is straightforward that this scalar signal $r(n)$ has the same cyclic and non-conjugate cyclic frequencies as the received signal ${\bf y}(n)$. Our goal is to find  filter ${\bf g}(z)$ producing a signal $r(n)$ that coincides with a filtered version of one of the source signals $(s_k)_{k=1, \ldots, K}$. This can be achieved by minimizing a cost function. In the following we investigate whether or not the Godard cost function is a good contrast function for mixtures containing BPSK signals. In a cyclostationary context and for a discrete time signal $r$, the Godard cost function is defined as

\begin{equation}
\label{eq:God}
J(r) = < \mathbb{E} \left( |r(n)|^{2} - 1 \right)^{2} > 
\end{equation} 
In order to express $J(r)$ in a more convenient way, we remark that
$r(n)$ can be written as 
\begin{equation}
\label{eq:expre-r(n)}
r(n) = \sum_{k=1}^{K} e^{2 i \pi n \delta f_k} [f_k(z)]s_k(n)
\end{equation}  
where $f_k(z)$ is the transfer function $f_k(z) = {\bf g}(z e^{-2 i \pi \delta f_k}) {\bf h}_k(z)$. We denote by $\|f_k\|$ the norm of filter $f_k(z)$ defined by 
$$||f_k||^{2} = \int_{-1/2}^{1/2} |f_k(e^{2 i \pi \nu})|^{2} S_{s_k}^{(0)}(e^{2 i \pi \nu}) \, d\nu$$
where $S_{s_k}^{(0)}(e^{2 i \pi \nu})$ represents the spectral density of signal $(s_k(n))_{n \in \mathbb{Z}}$. We finally define  
filter $\tilde{f}_k(z)$ and signal $\tilde{s}_k(n)$ by
\begin{equation}
\label{eq:def-tildef}
\tilde{f}_k(z) = \frac{f_k(z)}{\|f_k\|}, \; \tilde{s}_k(n) = [\tilde{f}_k(z)]s_k(n)
\end{equation}
If $\|f_k\| = 0$, we put $\tilde{f}_k(z) = 0$  and $\tilde{s}_k(n) = 0$. It is clear that $\| \tilde{f}_k \| = 1$, and that  
$<\mathbb{E}|\tilde{s}_k(n)|^{2}> = 1$. $r(n)$ can be written as
\begin{equation}
\label{eq:expre-r-bis}
r(n) = \sum_{k=1}^{K} \| f_k \| e^{2 i \pi n \delta f_k} \tilde{s}_k(n)
\end{equation}
and coincides with a filtered version of one of the source signal (up to the term $e^{2 i \pi n \delta f_k}$) if and only if the coefficients $(\|f_k\|)_{k=1, \ldots, K}$ satisfy 
$\|f_k\| = \delta(k - k_0) \|f_{k_0}\|$.
We state the following result 
\begin{proposition}
\label{prop:devlop-J}
The Godard cost function given by (\ref{eq:God}) can be expanded as
\begin{equation}
\label{eq:expre-J-developpee}
J(r) = \sum_{k=1}^{K} \beta(\tilde{s}_k) \|f_k\|^{4} +  \sum_{k_1 \neq k_2} l(\tilde{s}_{k_1}, \tilde{s}_{k_2}) \|f_{k_1}\|^{2}\|f_{k_2}\|^{2}
- 2 \sum_{k=1}^{K} \|f_k\|^{2} + 1
\end{equation} 
where $l(\tilde{s}_{k_1}, \tilde{s}_{k_2})$ and  $\beta(\tilde{s}_k)$ are defined respectively by 
\begin{equation}
\label{eq:expre-l}
 2 +  \mathrm{Re} \left[ 2 \sum_{\alpha \in I_{*}}   R_{\tilde{s}_{k_1}}^{(\alpha)}(0) \left(R_{\tilde{s}_{k_2}}^{(\alpha)}(0)\right)^{*}
+ \sum_{\alpha_c \in I_c}  R_{c,\tilde{s}_{k_1}}^{(\alpha_c - 2 \delta f_{k_1})}(0)  
\left(R_{c,\tilde{s}_{k_2}}^{(\alpha_c - 2 \delta f_{k_2})}(0)\right)^{*} \right]
\end{equation}
and by
\begin{equation}
\label{eq:def-beta}
 <c_4(\tilde{s}_k)> + 2 +  2   
\sum_{l=-1, 1} \left| R_{\tilde{s}_{k}}^{l \alpha_k}(0) \right|^{2} 
+  \sum_{l=-1,0, 1} \left| R_{c,\tilde{s}_{k}}^{l \alpha_k}(0) \right|^{2} 
\end{equation}
\end{proposition}{\bf Proof.} We start by writing $J(r)$ as
$$J(r) =  < \mathbb{E}|r(n)|^{4}> - 2 < \mathbb{E}|r(n)|^{2}> + 1$$
Using the relation
$$\mathbb{E}|r(n)|^{4} = c_4(r(n)) + 2  \left(\mathbb{E}|r(n)|^{2}\right)^{2} + \left| \mathbb{E}(r^{2}(n)) \right|^{2}$$
and the Parseval identities $< \left(\mathbb{E}|r(n)|^{2}\right)^{2} > = \sum_{\alpha \in I} |R_r^{(\alpha)}(0)|^{2}$ and 
$< \left|\mathbb{E}(r(n))^{2}\right|^{2} > = \sum_{\alpha \in I_c} |R_{c,r}^{(\alpha)}(0)|^{2}$, we immediately get that
\begin{equation}
\label{eq:expre-J-intermediaire}
J(r) =  < c_4(r(n))> + 2 \sum_{\alpha \in I} |R_{r}^{(\alpha)}(0)|^{2}  + \sum_{\alpha \in I_c} |R_{c,r}^{(\alpha_c)}(0)|^{2} - 2 R^{(0)}_r(0) + 1
\end{equation} 
Since $(\tilde{s}_k)_{k=1, \ldots, K}$ signals are independent we can write
\begin{eqnarray*}
<c_4(r(n))> &= & \sum_{k=1}^{K} \|f_k\|^{4} <c_4(\tilde{s}_k(n))> \\
R_{r}^{(\alpha)}(0) & = & \sum_{k=1}^{K}  \|f_k\|^{2} R_{\tilde{s}_k}^{(\alpha)}(0)
\end{eqnarray*}
where $\alpha$ represents one of the cyclic frequencies of $r$. 
For $\alpha = 0$, the last expression becomes $R_{r}^{(0)}(0) = \sum_{k=1}^{K}  \|f_k\|^{2}$ since we assumed that $<\mathbb{E}|\tilde{s}_k(n)|^{2}> =  R_{\tilde{s}_k}^{(0)}(0) = 1$. 
Furthermore, it is easily proved that if $\alpha_c$ is one of the non-conjugate cyclic frequencies of $r$, then the non-conjugate cyclic correlation coefficient of signal $e^{2 i \pi n \delta f_k} \tilde{s}_k(n)$ at $\alpha_c$ frequency and at time lag 0 coincides with $R_{c,\tilde{s}_k}^{(\alpha_c - 2 \delta f_k)}(0)$. This implies that : 
\[
R_{c,r}^{(\alpha_c)}(0)  =  \sum_{k=1}^{K}  \|f_k\|^{2} R_{c,\tilde{s}_k}^{(\alpha_c - 2 \delta f_k)}(0) 
\] 
Using these various expressions in (\ref{eq:expre-J-intermediaire}) we obtain the announced result. Notice that it is easy to establish that $\beta(\tilde{s}_k)$ is also given by
\begin{equation}
\label{eq:expre-beta}
\beta(\tilde{s}_k) = < \mathbb{E}|\tilde{s}_k(n)|^4>
\end{equation}
Note that, as shown in \cite{jal-che-lou-06-soumis}, $\beta(\tilde{s}_k) = < \mathbb{E}|\tilde{s}_k(n)|^4> \geq 1.$ 

Expression (\ref{eq:expre-J-developpee}) shows that $J(r)$ is a function of both the norms $(\|f_k\|^{2})_{k=1, \ldots, K}$,
and the unit norm filters $(\tilde{f}_k(z))_{k=1, \ldots K}$ defined by $\tilde{s}_k(n) = [\tilde{f}_k(z)]s_k(n)$, and
that these 2 sets of parameters are independent. Minimizing $J(r)$ with respect to ${\bf g}(z)$ is thus equivalent to minimizing 
 (\ref{eq:expre-J-developpee}) independently with respect to the norms $(\|f_k\|^{2})_{k=1, \ldots, K}$ and the unit norm filters $(\tilde{f}_k(z))_{k=1, \ldots K}$. 

In the following we study the minimization of $J(r)$ firstly when the different source signals do not have any non zero cyclic frequency in common nor any non-conjugate cyclic frequency in common and then we consider an opposite scenario where $K$  BPSK signals share the same baud rate and the same carrier frequency. 


\section{The source signals do not share the same cyclic and non conjugate cyclic frequencies}
\label{sec:differentes}

We first study the behavior of $J(r)$ when the source signals do not share the same cyclic and non conjugate cyclic frequencies. 
This situation is likely to occur when the different transmitters do not belong to the same network and it practically implies that 
$\forall k \neq l \in \{1 \ldots K \}$ $\alpha_k \neq  \alpha_l$ (i.e. $T_k \neq T_l$) and $\delta f_k \neq \delta f_l$ ($\Delta f_k \neq \Delta f_l$). 
In this context, the term $l(\tilde{s}_{k_1}, \tilde{s}_{k_2})$ reduces to the constant term 2, and $J(r)$ is given by
\begin{equation}
\label{eq:expre-J-differentes}
J(r) =  \sum_{k=1}^{K} \beta(\tilde{s}_k) \|f_k\|^{4} + 2 \sum_{k_1 \neq k_2} \|f_{k_1}\|^{2}\|f_{k_2}\|^{2}
- 2 \sum_{k=1}^{K} \|f_k\|^{2} + 1
\end{equation}
We now study the conditions under which the minimum of $J(r)$ is reached for a filter such that  $\|f_k\| = \delta(k - k_0) \|f_{k_0}\|$. For this, we follow \cite{jal-che-lou-06-soumis} and we first fix the unit norm filters $(\tilde{f}_k)_{k=1, \ldots, K}$ or equivalently the $(\beta(\tilde{s}_k))_{k=1, \ldots, K}$  coefficients. Then, we consider the problem of minimizing $J$ with respect only to the $(\|f_k\|^2)_{k=1, \ldots, K}$. This is an easy task because, as a function of the  $(\|f_k\|^2)_{k=1, \ldots, K}$ norms, $J(r)$ has a simple expression which allows the following result to be derived  
\begin{theorem}
\label{th:differentes}
The minimum of $J(r)$ w.r.t.  $(\|f_k\|^{2})_{k=1, \ldots, K}$ is reached for sequences such that 
$\|f_k \|^{2} = \delta(k-k_0) \|f_{k_0} \|^{2}$ for a certain $k_0$ index if and only if
\[
\min_{k=1, \ldots, K} \beta(\tilde{s}_k) < 2
\]
and if this minimum is reached for the index $k_0$. Moreover, the minimum value of $J$ is equal to $1 - \frac{1}{\beta_{\text{min},k_0}}$.
\end{theorem}
\begin{corollary}
If the sources do not share the same cyclic and non conjugate cyclic frequencies, the global minimization of the Godard cost function allows to extract all the source signals using a deflation approach if 
\begin{equation}
\label{eq:condition-differentes}
\beta_\text{min,k} =  \min_{\tilde{f}_k, \|\tilde{f}_k\|=1} \beta(\tilde{s}_k) < 2, \; \mbox{for each $k=1, \ldots, K$}
\end{equation}
\end{corollary}
The proof of this theorem can be found in \cite{jal-che-lou-06-soumis}.
It remains to check if condition (\ref{eq:condition-differentes}) holds. For circular linearly modulated signals, (\ref{eq:condition-differentes}) has been analytically proved in \cite{jal-che-lou-06-soumis}. In the case of BPSK signals, the following result can be proved using a similar approach.

\begin{figure}[!h]
  \begin{center}
         \subfigure[]{\label{fig:betamin_bpsk}\includegraphics[scale=0.45]{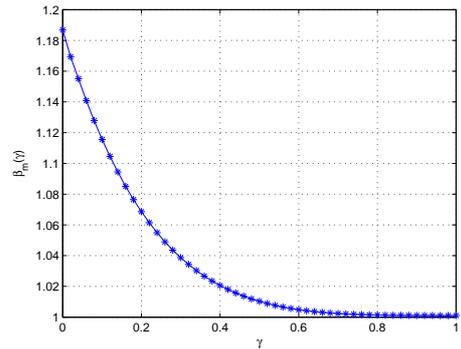}}  
         \subfigure[]{\label{fig:betamin_circulaire}\includegraphics[scale=0.45]{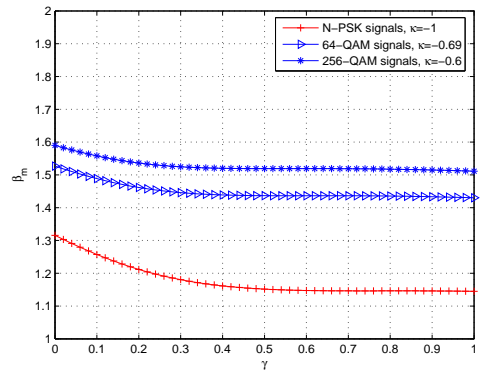}}
   \end{center}
   \caption{$\beta_{\text{min}}$ as a function of $\gamma$ in the case of (a) BPSK signals and (b) cicular signals.}
   \label{fig:beta_mins}
\end{figure}

\begin{proposition}
\label{prop:betaminBPSK}
Consider a BPSK signal with symbol period $T$ and excess bandwidth $0 < \gamma < 1$, and assume that the sampling period $T_e$ does not belong to 
$\{ T, \frac{T}{2}, \frac{T}{3}, \frac{2T}{3}\}$. Denote by $\kappa$ the kurtosis of the corresponding binary symbol sequence, $\kappa = - 2$. Then, $\beta_{\text{min}} = \min_{\tilde{f}, \|\tilde{f}\|=1} \beta([\tilde{f}(z)]s(n))$ 
is given by 
\begin{equation}
\label{eq:expre-betamin}
\beta_{\text{min}}   =  \inf_{f_a \in {\cal F}( [-\frac{1+\gamma}{2T}, \frac{1+\gamma}{2T}])} \, \Phi(f_a)
\end{equation}
where $\Phi(f_a)$ is defined by 
\begin{equation}
\label{eq:expre-Phi}
\begin{array}{c} 
\Phi(f_a) =  \kappa T  \frac{ \int_{\mathbb{R}} |f_{a}(t)|^4 dt}{(\int_{\mathbb{R}} |f_{a}(t)|^2 dt)^2} + 2 + 4 \left| \frac{\int_{\mathbb{R}} |f_{a}(t)|^2 e^{-2i\pi \frac{t}{T}} dt}{\int_{\mathbb{R}} |f_{a}(t)|^2 dt}\right|^2 \nonumber \\
 + \frac{\left| \int_{\mathbb{R}} f_{a}(t)^2  dt \right|^{2} }{\left(\int_{\mathbb{R}} |f_{a}(t)|^2 dt \right)^2} + \frac{\left| \int_{\mathbb{R}} f_{a}(t)^2 e^{-2i\pi \frac{t}{T}} dt \right|^{2}}{\left( \int_{\mathbb{R}} |f_{a}(t)|^2 dt \right)^2} +  \frac{\left| \int_{\mathbb{R}} f_{a}(t)^2 e^{2i\pi \frac{t}{T}} dt \right|^{2}}{\left( \int_{\mathbb{R}} |f_{a}(t)|^2 dt \right)^2} 
\end{array}
\end{equation}
Moreover, if we define $\eta_{\text{min}}$ by $\eta_{\text{min}} = \min_{\|\tilde{f}\|=1} < c_4(\tilde{s})>$, 
then 
\begin{equation}
\label{eq:expre-etamin}
\eta_{\text{min}}   =  \inf_{f_a \in {\cal F}( [-\frac{1+\gamma}{2T}, \frac{1+\gamma}{2T}])} \,  \kappa T  \frac{ \int_{\mathbb{R}} |f_{a}(t)|^4 dt}{(\int_{\mathbb{R}} |f_{a}(t)|^2 dt)^2}
\end{equation}   
\end{proposition} 
We give the proof of this result in the  \ref{ap:A}.
\begin{remarque}
\label{rem:cas-particuliers}
If $T_e \in \{ T, \frac{T}{2}, \frac{T}{3}, \frac{2T}{3} \}$, the expression and, as a consequence, the value of  $\beta_{\text{min}}$, 
is different from (\ref{eq:expre-betamin}). $\beta_{\text{min}}$, as a function of $T_e$, is therefore a constant function 
except in such points as $ \{ T, \frac{T}{2}, \frac{T}{3}, \frac{2T}{3} \}$ where it has a different value. We are therefore dealing with a discontinuous function. In order to illustrate this point, we consider as an example the case where $T_e = T$. For this sampling rhythm to satisfy the condition of Shannon it is necessary and sufficient that the excess bandwidth factor be $0$. 

Under these conditions, it is well known that $\beta_{\text{min}}$ equals 1 and $\eta_{\text{min}} = -2$, while we will soon see that $\beta_{\text{min}} \simeq 1.19$ and $\eta_{\text{min}} = -1.36$ if $T_e$ does not belong to $\{ T, \frac{T}{2}, \frac{T}{3}, \frac{2T}{3} \}$. For simplicity reasons, we prefer not to give the expressions of $\beta_{\text{min}}$ if $T_e \in \{ \frac{T}{2}, \frac{T}{3}, \frac{2T}{3} \}$. In any case, the probability of $T_e$ being equal to one of these values is obviously null in a blind context. For this reason we suppose in the following that $T_e$ does not belong to $\{ T, \frac{T}{2}, \frac{T}{3}, \frac{2T}{3} \}$. 
\end{remarque}
As  ${\cal F}( [-\frac{1+\gamma_1}{2T}, \frac{1+\gamma_1}{2T}]) \subset {\cal F}( [-\frac{1+\gamma_2}{2T}, \frac{1+\gamma_2}{2T}])$ if $\gamma_1 < \gamma_2$, (\ref{eq:expre-betamin}) implies that considered as a function of $\gamma$, $\beta_{\text{min}}(\gamma)$ is decreasing. This observation allows us to make the following statement :
\begin{proposition}
\label{prop:decroissance-betamin}
Function $\gamma \rightarrow \beta_{\text{min}}(\gamma)$ is decreasing when $\gamma$ varies from $0$ to $1$. Consequently, 
$\beta_{\text{min}}(\gamma)$ is strictly inferior to 2 for all $\gamma$ if and only if 
$\beta_{\text{min}}(0) < 2$. 
\end{proposition}
The main interest of proposition \ref{prop:decroissance-betamin} is that if a function $f_a(t)\in{\cal F}([-\frac{1}{2T},\frac{1}{2T}])$ (corresponding to $\gamma = 0$), then the integrals
\[
\int_{\mathbb{R}} |f_{a}(t)|^2 e^{-2i\pi \frac{t}{T}} dt, \;  \int_{\mathbb{R}} f_{a}(t)^2 e^{-2i\pi \frac{t}{T}} dt, \;  \int_{\mathbb{R}} f_{a}(t)^2 e^{2i\pi \frac{t}{T}} dt
\]
vanish. This result is a direct application of the inequality of Parseval. 
The expression of $\beta_{\text{min}}(0)$ is therefore 
\begin{equation}
\label{eq:beta-min-zero}
\beta_{\text{min}}(0) = \min_{f_a \in \mathcal{F}([-\frac{1}{2T}, \frac{1}{2T}])} \kappa T  \frac{ \int_{\mathbb{R}} |f_{a}(t)|^4 dt}{(\int_{\mathbb{R}} |f_{a}(t)|^2 dt)^2} + 2 +
 \frac{\left| \int_{\mathbb{R}} f_{a}(t)^2  dt \right|^{2} }{\left(\int_{\mathbb{R}} |f_{a}(t)|^2 dt \right)^2}
\end{equation}
It is easy to notice that $\beta_{\text{min}}$ does not depend of $T$ and that the theoretical expressions (\ref{eq:expre-betamin}) and  (\ref{eq:expre-etamin}) of $\beta_{\text{min}}$ and $\eta_{\text{min}}$ can be used in order to compute the numerical values of these functions for all the values of $\gamma \in [0,1]$ via the approach proposed in \cite{jal-che-lou-06-soumis}.
Figure \ref{fig:betamin_bpsk} gives a numerical representation of $\beta_{\text{min}}$ as a function of $\gamma$ in the case of BPSK signals. Moreover, we have found that $\eta_{\text{min}} \simeq 0.68 \kappa (1+\gamma)$ and is equal to $ -1.36 (1+\gamma)$ in the case of BPSK signals, since $\kappa=-2$.
Figure \ref{fig:betamin_bpsk} also confirms the decreasing nature of  $\beta_{\text{min}}$ with respect to $\gamma$, and  the fact that for a BPSK modulated signal
$\beta_{\text{min}} < 2$ for all $\gamma$  provided that $T_e$ does not belong to $\{ T, T/2, T/3, 2T/3 \}$. If $T_e=T$, as we have already mentioned,  $\beta_{\text{min}} = 1$; if $T_e$ equals one of the other possible values, we can directly verify that $\beta_{\text{min}}$ remains strictly inferior to 2. We can therefore enunciate the following result:

\begin{proposition}
\label{prop:separationOK}
In the case of circular or BPSK transmitted signals, not sharing any non zero cyclic frequency nor any non conjugate cyclic frequency, the minimization of the constant modulus criterion, along with a deflation approach allows the extraction of all sources.
\end{proposition}
\begin{remarque}
\label{rem:ordre-d'extraction}
Notice that the values of $\beta_{\text{min}}$ for a BPSK modulated signal are smaller than the ones we observe for linearly modulated circular signals which we represent in figure \ref{fig:betamin_circulaire}. This means that if a BPSK modulated signal is mixed with circular modulated signals, the BPSK source will very often be the first one extracted when using a deflation approach. 
\end{remarque}

Despite its undeniable importance, proposition \ref{rem:ordre-d'extraction} is not completely convincing as to the pertinence of the proposed approach. In practice, the search for filter ${\bf g}(z) = \sum_{l=-L}^{L} {\bf g}(l) z^{-l}$  which extracts a source from the mixture is done by minimizing an estimator $\hat{J}(r)$ of $J(r)$. Furthermore, the minimization of $\hat{J}(r)$ is carried out by means of iterative algorithms such as the steepest descent or Newton algorithms who are not guaranteed to converge toward the global minimum of $\hat{J}$ and may very well converge toward a {\bf local minimum} instead. It is therefore necessary to verify that ${J}$ does not have any {\bf non separating} local minima.  Under a technical assumption, the following result can be established
\begin{proposition}
\label{prop:minlocaux-differents}
Assume that at least one of the functions $\tilde{f}_k \rightarrow \beta([\tilde{f}_k(z)]s_k(n))$ defined on the set of all unit norm filters has no local minimum 
$\tilde{f}_k^{*}$ such that $\beta([\tilde{f}^{*}_k(z)]s_k(n)) \geq 2$. Then, the argument of each local minimum of the Godard cost function is a separating 
filter. 
\end{proposition}
{\bf Proof.} We define the following quantities $u = (\sum_{k=1}^{K} \|f_k\|^{2})^{1/2}$ and $v_k = \frac{\|f_k\|}{u}$. Expression (\ref{eq:expre-J-differentes}) of $J(r)$ then becomes 
$$
J(r) = u^4\left[\sum_{k=1}^{K}{\beta({\tilde{s}_k})v_k^4} + 2\sum_{k_1\neq k_2}{v_{k,1}^2v_{k,2}^2}\right] - 2 u^2\left(\sum_{k=1}^{K} v_k^{2}\right) +1
$$ 
In the following we pose ${\mathbf{\beta}} = (\beta(\tilde{s}_{1}), \ldots, \beta(\tilde{s}_{K}))^{T}$ and we denote by $T(\bf{v},\bs{\beta})$ the expression multiplying the term $u^4$. It is clear that $\sum_{k=1}^{K} v_k^{2} = 1$. Since $ \sum_{k_1\neq k_2}{v_{k,1}^2v_{k,2}^2} = (\sum_{k=1}^{K}{v_{k}^2})^2 - \sum_{k=1}^{K} v_k^{4}$ we obtain a simpler expression for $T(\bf{v},\bs{\beta})$ 
$$ 
T({\bf v},{\bs\beta}) = 2 + \sum_{k=1}^{K}{v_k^4(\beta(\tilde{s}_k)-2)}
$$
$J(r)$ is thus given by:
$$
J(r) = u^{4} T({\bf v}, {\bs \beta}) - 2 u^{2} + 1
$$ 
We consider a local minimum $(f_1^{*}(z), \ldots, f_K^{*}(z))^{T}$ of $J(r)$, and denote by  $u_*$, ${\bf v}_{*}$, $\tilde{f}_{k}^{*}$, $\tilde{s}_k^{*}, {\bs \beta}_{*}$ the corresponding values of $u, {\bf v}, \tilde{f}_k, \tilde{s}_k, {\bs \beta}$. It is easy to check that 
the point ${\bf v}_{*}$ is a local minimum of the function ${\bf v} \rightarrow T({\bf v}, {\bs \beta}_*)$. As at least one the coefficients  
$(\beta(\tilde{s}_k^{*}) - 2)$ is strictly negative,  $v_{k}^{*} = \delta(k-k_0) v_{k_{0}}^{*}$ where $k_0$ is one of the index for which  $\beta_{k_{0},*} - 2 < 0$ 
(see e.g. \cite{r-del-lou-95}). This implies that $\|f_{k,*}\| =  \delta(k-k_0) \|f_{k_{0}}^{*} \|$, and that the local minimum $f_{1,*}(z), \ldots, f_{K,*}(z)$ is a 
separating filter. 
It is difficult to check analytically whether or not it exists $k$ for which $\tilde{f}_k \rightarrow \beta([\tilde{f}_k(z)]s_k(n))$ has no local minimum 
$\tilde{f}_k^{*}$ such that $\beta([\tilde{f}^{*}_k(z)]s_k(n)) \geq 2$. However, this condition probably holds because the steepest descent  minimization 
algorithms of the functions  $\tilde{f}_k \rightarrow \beta([\tilde{f}_k(z)]s_k(n))$ we have run always converge toward a point for which $\beta([\tilde{f}_k(z)]s_k(n)) < 2$.

In sum, the above results indicate that if the source signals do not share the same cyclic and non conjugate cyclic frequencies, then, the minimization of the
Godard cost function allows to extract circular and BPSK source signals. In this context, it is therefore possible to separate the source signals
without any knowledge of their cyclic and non conjugate cyclic frequencies. 

\section{K BPSK sources sharing the same baud-rate and the same carrier frequency}
\label{sec:egales}

In this section, we consider the opposite situation, when all the source signals are BPSK signals with the same 
baud rate $T$, the same carrier frequency offset $\Delta f$, and the same excess bandwidth $\gamma$. We also denote by $\alpha$ and $\delta f$ the terms 
$\alpha = T_e/T$ and $\delta f = \Delta f T_e$. Recall that the sampling rate $T_e$ is assumed not to belong to $\{ T, T/2, T/3, 2T/3 \}$.

\subsection{Existence of spurious local minima for $K=2$ and $\gamma=0$}
\begin{figure}[!h]
  \begin{center}
         \subfigure[] {\label{fig:histogramme-valeurs-J}\includegraphics[scale=0.45]{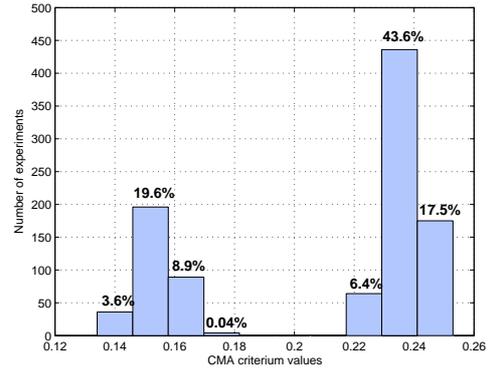}}  
         \subfigure[]{\label{fig:histogramme-SINR}\includegraphics[scale=0.45]{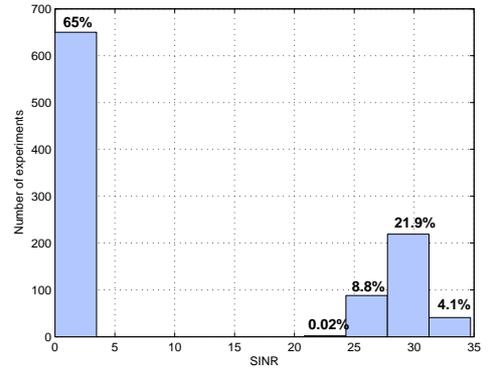}}
   \end{center}
   \caption{Histograms of (a) $\hat{J}(r)$ values and (b)SINR values, obtained after extracting a source from a mixture of 2 identical BPSK signal with $\gamma = 0$.}
   \label{fig:histogramme-SINR}
\end{figure}

Our purpose is to support the conjecture that the Godard cost function has  non separating local minima, and that the minimization algorithms often converge toward these spurious points. In order to justify this, we assume that the common excess bandwidth $\gamma$ of the 2 source signals is equal to 0. In this context, the cyclic and non conjugate cyclic correlations coefficients at frequencies $\pm \alpha$ are zero. Expression 
(\ref{eq:expre-J-developpee}) of $J(r)$ thus reduces to 
\begin{align}
\label{eq:expre-J-gamma=0}
&J(r) = \beta(\tilde{s}_1) \|f_1\|^{4} + \beta(\tilde{s}_2) \|f_2\|^{4} + \\
&2 \|f_1\|^{2} \|f_2\|^{2} \left(2 + \mathrm{Re}(R^{(0)}_{c, \tilde{s}_1}(0) 
R^{(0)}_{c, \tilde{s}_2}(0)^{*}) \right) - 2 \left( \|f_1\|^{2} + \|f_2\|^{2} \right) + 1\nonumber
\end{align}
where $\beta(\tilde{s}_i)$ is given by 
\[
\beta(\tilde{s}_i) = <c_4(\tilde{s}_i)> + 2 + |R^{(0)}_{c, \tilde{s}_i}(0)|^{2}
\]
for $i=1,2$. This expression is formally similar to the one of $J$ in the case where the
2 sources are circular with a non zero excess bandwidth (see \cite{jal-che-lou-06-soumis}), except  
that the cyclic correlation coefficients $R^{(0)}_{c, \tilde{s}_i}(0)$ are replaced by $2 R^{(\alpha)}_{\tilde{s}_i}(0)$. 
An analog of the condition $|2 R^{(\alpha)}_{\tilde{s}_i}(0)| \leq 1$, which plays an important role in \cite{jal-che-lou-06-soumis}, can also be proved true for the cyclic correlation coefficients $R^{(0)}_{c, \tilde{s}_i}(0)$, i.e.  $|R^{(0)}_{c, \tilde{s}_i}(0)| \leq 1$. Considering the definition
 of $\tilde{s}_i$ in (\ref{eq:def-tildef}), we can write
$$
R^{(0)}_{c, \tilde{s}_k}(0) = \int_{\mathbb{R}} \hat{\tilde{f}}(e^{2 i\pi\nu }) \hat{\tilde{f}}(e^{-2 i\pi\nu })  S_{c,s_k}^{(0)}(e^{2 i\pi\nu }) \, d\nu
$$
As signal $s_k$ is real valued, ${S}_{c,s_k}^{(0)}$ coincides with the spectrum ${S}_{s_k}^{(0)}$ of $s_k$, and is an even function. Using the Schwartz inequality, we get immediately that  $|R^{(0)}_{c, \tilde{s}_i}(0)| \leq 1$. It is therefore possible to use Theorem 2 of \cite{jal-che-lou-06-soumis} 
established in the circular case to prove that if $\beta_{\text{min}}$ and $\eta_{\text{min}}$ defined in Proposition \ref{prop:betaminBPSK} verify 
\begin{equation}
\label{ameliore:cond1}
\begin{cases}
& -3 \beta_{{\text{min}}}+5+  \eta_{{\text{min}}}  >  0 \\
& 2(\beta_{{\text{min}}}-1)\beta_{{\text{min}}}-4 \left(1-\frac{1}{2}\sqrt{2(\beta_{{\text{min}}}-1)-1-  \eta_{{\text{min}}}}\right)^{2} < 0 
\end{cases}
\end{equation} 
then, the argument of the global minimum of $J(r)$ is a separating filter, and the minimum value of $J(r)$ coincides with 
$1 - 1/\beta_{\text{min}}$. For $\gamma = 0$, $\beta_{\text{min}} \simeq 1.19$, $\eta_{\text{min}} \simeq -1.36$, and it is easily checked that the 2 conditions above
are satisfied. The global minimization of $J(r)$ therefore allows to separate the 2 BPSK signals. Moreover, 
$1 - 1/\beta_{\text{min}} \simeq 0.16$. However, $J(r)$ may have non separating local minima, toward which a steepest descent minimization algorithm of $J(r)$ often converges. In order to define these local minima, we denote by $\tilde{f}_1^{*}(z)$ 
one of the arguments of the global minimum of $\beta([\tilde{f}_1(z)]s_1(n))$ over the set of unit norm filters 
{\bf with real coefficients}. We denote by $\beta_{1,min}$ the corresponding minimum. It is easy to show that  
$\beta_{1,min}$ can be evaluated using Proposition \ref{prop:betaminBPSK}, by minimizing the function $\Phi_a$ over {\bf the real elements} of ${\cal F}([-1/2T, 1/2T])$ when $\gamma = 0$. In these conditions, it can be shown that $\beta_{1,min}$ coincides with $\eta_{\text{min}} + 3$, i.e. that $\beta_{1,min} \simeq  1.64$. We now consider the unit norm filter with imaginary coefficients $\tilde{f}_2^{*}(z) = i \tilde{f}_1^{*}(z)$. It is clear
that $\beta([\tilde{f}_2^{*}(z)]s_2(n))$ coincides with $\beta_{1,min}$. We finally define filters 
$f_i^{*}(z)$ for $i=1,2$ by 
\begin{equation}
\label{eq:def-s-tilde}
f_i^{*}(z) = \frac{1}{(1+\beta_{1,min})^{1/2}} \tilde{f}_i^{*}(z)
\end{equation}
If $r_{*}(n) = [f_1^{*}(z)]s_1(n) +  [f_2^{*}(z)]s_2(n)$, one can check that 
$J(r_{*}) = 1 - 2/(1+\beta_{1,min}) \simeq 0.25$. Although we have not been able to analytically prove these non separating points to be a local minimum of $J$, we have observed that the steepest descent minimization algorithm of $J(r)$ very often converges to one of these points rather than toward the argument of the separating global minimum of $J$. To verify this, we present in Figure \ref{fig:histogramme-valeurs-J} an histogram of the values of $J(r)$ at convergence of the steepest descent minimization algorithm.
We used 1000 experiments, each corresponding to different randomly selected propagation channels, and we assumed the thermal noise to be negligible. The figure clearly shows that in more than half of the experiments the final value of $\hat{J}(r)$ corresponds to $1 - \frac{1}{1.32} \simeq 0.25$ which is associated to a local minima rather than to the value of the global minimum of $J$ which is $1 - \frac{1}{\beta_{\text{min}}} = 1 - \frac{1}{1.19} \simeq 0.16$. 

In order to verify that the value $1 - \frac{1}{1.32}$  does not correspond to a separating filter, we present in figure \ref{fig:histogramme-SINR} an histogram of the signal to interference and noise ratio (SINR) associated to the filters determined by minimizing $\hat{J}(r)$. We define the SINR as the ratio between the power of  signal $r_1$, representing the contribution of the extracted signal filtered by the extracting filter and the power of signal $r_2$  which represents the contribution of the other transmitted signal filtered by the same filter. It is clear that if the filter is perfectly adjusted then the SINR must equal $+\infty$ in the absence of thermal noise. 
The experiments we presented thus tend to confirm the fact that $J(r)$ has non separating local minima and that the steepest descent algorithm converges very often toward one of them.
\subsection{A new cost function}

A simple modification of the Godard cost function allows to overcome the aforementioned problems, provided that the most significant non-conjugate cyclic frequencies of the received signal are known or can be correctly estimated by the receiver. We recall that for a mixture of BPSK modulated signals sharing the same carrier frequency, the most significant cyclic frequency is $2\delta f$.  

In the following, we assume that the carrier frequency offset $\delta f$ is known or correctly estimated at the receiver side, and consider the cost function $J^{'}(r)$ defined by 
\begin{align}
\label{eq:def-J'}
J^{'}(r)& = J(r) - |R^{(2 \delta f)}_{c,r}(0)|^{2} \\
& = < \mathbb{E} \left(|r(n)|^{2} - 1 \right)^{2} > - \left| < \mathbb{E}(r^{2}(n)) e^{-2 i \pi n 2 \delta f}> \right|^{2}\nonumber
\end{align}

$J^{'}(r)$ is obtained by subtracting from $J(r)$ the modulus square of the non conjugate cyclic correlation coefficient at time lag 0 and at non conjugate cyclic frequency $2 \delta f$. Using the expression of $J(r)$, we immediately obtain that 
\begin{equation}
\label{eq:expre-J'-developpee}
J^{'}(r) = \sum_{k=1}^{K} \beta^{'}(\tilde{s}_k) \|f_k\|^{4} +  \sum_{k_1 \neq  k_2} l^{'}(\tilde{s}_{k_1}, \tilde{s}_{k_2}) \|f_{k_1}\|^{2}\|f_{k_2}\|^{2}
- 2 \sum_{k=1}^{K} \|f_k\|^{2} + 1
\end{equation} 
where the term $l^{'}(\tilde{s}_{k_1}, \tilde{s}_{k_2})$ is given by  
\begin{equation}
\label{eq:expre-l'}
2 +  \mathrm{Re} \left[ 2 \sum_{l=-1,1}  R_{\tilde{s}_{k_1}}^{(l \alpha)}(0) \left(R_{\tilde{s}_{k_2}}^{(l \alpha)}(0)\right)^{*} 
+ \sum_{l=-1, 1} R^{ (l \alpha)}_{c, \tilde{s}_{k_1}}(0) R^{(l \alpha)}_{c,\tilde{s}_{k_2}}(0)^{*} \right]     
\end{equation}
and where $\beta^{'}(\tilde{s}_k)$ is defined by 
\begin{equation}
\label{eq:def-beta'-BPSK}
 <c_4(\tilde{s}_k)> + 2 +  2 \sum_{l = -1, 1}   
\left| R_{\tilde{s}_{k}}^{l \alpha}(0) \right|^{2}
+ \sum_{l = -1, 1}  \left| R_{c,\tilde{s}_{k}}^{(l \alpha)}(0) \right|^{2} 
\end{equation}
$\beta^{'}(\tilde{s}_k)$ also equals
\begin{equation}
\label{eq:def-bis-beta'-BPSK}
\beta^{'}(\tilde{s}_k) = \beta(\tilde{s}_k) - \left| R_{c,\tilde{s}_{k}}^{(0)}(0) \right|^{2} 
\end{equation}
In order to give some insight on $J^{'}$, we first consider the case $\gamma=0$. Expression (\ref{eq:expre-J'-developpee}) of $J^{'}(r)$  therefore becomes
\[
J^{'}(r) = \sum_{k=1}^{K} \beta^{'}(\tilde{s}_k) \|f_k\|^{4} + 2 \sum_{k_1 \neq k_2} \|f_{k_1}\|^{2}  \|f_{k_2}\|^{2}
- 2 \sum_{k=1}^{K} \|f_k\|^{2} + 1
\]
Furthermore, $\beta^{'}(\tilde{s}_i)$ now equals $\beta^{'}(\tilde{s}_i) = <c_4(\tilde{s}_i)> + 2$. The expression of $J^{'}(r)$ is thus similar to 
(\ref{eq:expre-J-differentes}), except that $\beta(\tilde{s}_i)$ is now replaced by $\beta^{'}(\tilde{s}_i)$. It is easy to check that
$ <c_4(\tilde{s}_i)> < 0$, so that $\beta^{'}(\tilde{s}_i) < 2$ for each $i$. Theorem \ref{th:differentes} and Proposition \ref{prop:minlocaux-differents}
thus imply that the global minimum and the local minima of $J^{'}$ are separating filters. This shows that the minimization of $J^{'}(r)$ allows to 
separate the $K$ BPSK signals if $\gamma = 0$. 

In order to extend this result to the more general case where $\gamma > 0$, we now show that the argument of the minimum value of  $J'(r)$ corresponds to a separating filter.
Contrary to the case where the transmitted signals all had different cyclic frequencies and different non conjugate cyclic frequencies, it is no longer possible to directly characterize the global minimum of $J^{'}(r)$ since its analytical form is too complex. We overcome this difficulty by  using the following result stated and proved in 
\cite{jal-che-lou-06-soumis}:
\begin{proposition}
\label{prop:minoration}
Let $m(r)$ be a positive function such that for any filtered version $r(n)=[\mathbf{f}(z)]s(n)$ we have
$$J^{'}(r) \geq m(r)$$
Assume that the infimum of $m(r)$ is reached if and only if signal $r(n)$ coincides with a filtered version of one of the source signals. Let $r_{*}(n) = [f_{k_0,*}(z)]s_{k_0}(n)$ be one of the signals for which $\inf_{\mathbf{f}(z)}{m(r)}=m(r_{*})$. If $m(r_{*})=J^{'}(r_{*})$, then
$$\inf_{\mathbf{f}(z)}{J^{'}(r)}=J^{'}(r_{*}) $$
and the infimum is reached if and only if $r(n)$ coincides with one of the $r_{*}$ specified above.
\end{proposition}

In order to derive a function $m(r)$ satisfying the conditions of Proposition \ref{prop:minoration}, we prove the following result.

%
\begin{proposition}
\label{prop:minoration-l'}
The following inequality holds:
\begin{equation}
\label{eq:minoration-l'}
  \mathrm{Re}  \left[ 2 \sum_{l=-1,1}   R_{\tilde{s}_{k_1}}^{(l \alpha)}(0) \left(R_{\tilde{s}_{k_2}}^{(l \alpha)}(0)\right)^{*}
+ \sum_{l=-1, 1} R^{ (l \alpha)}_{c, \tilde{s}_{k_1}}(0) R^{(l \alpha)}_{c,\tilde{s}_{k_2}}(0)^{*} \right]  \geq -3/2
\end{equation}
\end{proposition}
We give the proof of this result in \ref{ap:B}.
Consider the function $m(r)$ defined by 
\begin{equation}
\label{eq:def-m}
m(r) = \beta^{'}_{\text{min}} \left( \sum_{k=1}^{K} \|f_k\|^{4} \right) + \frac{1}{2} \left( \sum_{k_1 \neq k_2} 
 \|f_{k_1}\|^{2}\|f_{k_2}\|^{2} \right) - 2 \sum_{k=1}^{K} \|f_k\|^{2} + 1
\end{equation}
where we denote by $\beta^{'}_{\text{min}}$ the quantity 
\[
\beta^{'}_{\text{min}} = \beta^{'}_{min,k}
\]
with $\beta^{'}_{min,k} = \min_{\|\tilde{f}_k\| = 1} \beta^{'}(\tilde{s}_k)$. Recall that the signals present in the analysed bandwidth are of the same nature and therefore all $(\beta^{'}_{min,k})_{k=1, \ldots, K}$ are equal. 
Since relation (\ref{eq:minoration-l'}) is verified, it is clear that $l^{'}(\tilde{s}_{k_1}, \tilde{s}_{k_2}) \geq 1/2$. Moreover, the 
$(\beta^{'}(\tilde{s}_k))_{k=1, \ldots, K}$ are all greater than $\beta^{'}_{\text{min}}$. This implies that for all $r$, $J^{'}(r) \geq m(r)$. 
We show that if $\beta^{'}_{\text{min}} < 1/2$, then, the global minimum of $m(r)$ is reached if all $(\|f_k\|)_{k=1, \ldots, K}$ are null except for 1, i.e. if $r(n)$ coincides with a filtered version of one of the sources. In order to establish this result, we pose $u^{2} = \sum_{k=1}^{K} \|f_k\|^{2}$, $v_k = \frac{\|f_k\|}{u}$, ${\bf v} = (v_1, \ldots, v_K)^{T}$, and we define $t({\bf v})$ as
\[
t({\bf v}) = (\beta^{'}_{\text{min}} - \frac{1}{2}) \sum_{k=1}^{K} v_k^{4} + \frac{1}{2}
\]
It is easy to verify that 
\[
m(r) = u^{4} t({\bf v}) - 2 u^{2} + 1
\]
and that the global minimum of $m(r)$ is reached in a point $(u_{*}, {\bf v}_{*})$ for which $t({\bf v}_*)$ is minimum 
and $u_*^{2} = \frac{1}{t({\bf v}_*)}$. The value of this minimum is then $1 - \frac{1}{t({\bf v}_*)}$. To conclude it suffices to remark that if $\beta^{'}_{\text{min}} - 1/2 < 0$, then the minimum of $t({\bf v})$ is reached if and only if all the components of ${\bf v}$ are null except for one who is equal to $1$, which corresponds to all $\|f_k\|$ being null except for one of them (\cite{r-del-lou-95}). Furthermore $t({\bf v}_*)$ is equal to   $\beta^{'}_{\text{min}}$, $u^{2}_{*} = \frac{1}{\beta^{'}_{\text{min}}}$ and the minimum value of  $m(r)$ is $1 - \frac{1}{\beta^{'}_{\text{min}}}$. 
In the following we denote by $k_0$ one of the index for which $\beta^{'}_{\text{min}} = \beta^{'}_{min, k_0}$, and by $\tilde{f}_{k_0,*}$ 
a unit norm filter for which $\beta^{'}_{min,k_0} = \beta^{'}([\tilde{f}_{k_0,*}(z)]s_{k_0}(n))$, and we pose $f_{k_0,*}(z) = u_{*} \tilde{f}_{k_0,*}(z)$. The minimum of $m(r)$ is reached if $r_*(n) = [f_{k_0,*}(z)]s_{k_0}(n)$, and  $J^{'}(r_*)$ coincides with 
$m(r_*) = 1 - \frac{1}{\beta^{'}_{\text{min}}}$. Proposition \ref{prop:minoration} then states that the global minimum of $J^{'}$ is reached only if $r(n)$ is a filtered version of $s_{k_0}(n)$. We have thus established the following result 
\begin{proposition}
\label{prop:separation-beta<1/2}
If $\beta^{'}_{\text{min}} < 1/2$, then the minimization of $J^{'}(r)$ allows the extraction of one of the sources from the mixture. 
\end{proposition}
We must now verify whether the condition $\beta^{'}_{\text{min}} < 1/2$ is satisfied or not. Following the same reasoning as in the case of $\beta_{\text{min}}$,  we can easily adapt proposition \ref{prop:betaminBPSK} by simply replacing the expression (\ref{eq:expre-betamin}) with
\begin{equation}
\label{eq:expre-betamin-prim}
\beta^{'}_{\text{min}}   =  \inf_{f_a \in {\cal F}( [-\frac{1+\gamma}{2T}, \frac{1+\gamma}{2T}])} \, \Phi^{'}(f_a)
\end{equation}
where $\Phi^{'}(f_a)$ is defined as 
\begin{align}
\label{eq:expre-Phi-prim} 
\Phi^{'}(f_a) = & \kappa T  \frac{ \int_{\mathbb{R}} |f_{a}(t)|^4 dt}{(\int_{\mathbb{R}} |f_{a}(t)|^2 dt)^2} + 2 + 4 \left(\frac{\int_{\mathbb{R}} |f_{a}(t)|^2 e^{-2i\pi \frac{t}{T}} dt}{\int_{\mathbb{R}} |f_{a}(t)|^2 dt}\right)^2 \nonumber \\
& + \frac{\left| \int_{\mathbb{R}} f_{a}(t)^2 e^{-2i\pi \frac{t}{T}} dt \right|^{2}}{\left( \int_{\mathbb{R}} |f_{a}(t)|^2 dt \right)^2} +  \frac{\left| \int_{\mathbb{R}} f_{a}(t)^2 e^{2i\pi \frac{t}{T}} dt \right|^{2}}{\left( \int_{\mathbb{R}} |f_{a}(t)|^2 dt \right)^2} 
\end{align}

The expression of $\Phi^{'}(f_a)$  is obtained directly by subtracting from the expression of $\Phi(f_a)$ (\ref{eq:expre-Phi}) the term due to the square modulus of the non conjugate cyclic coefficient of $s(n)$ at the non conjugate cyclic frequency $0$. As in the case of $\beta_{\text{min}}$, this result implies that  $\beta^{'}_{\text{min}}$ is a decreasing function of the excess bandwidth factor $\gamma$. We can thus formulate the following statement:

\begin{proposition}
\label{prop:decroissance-betamin-prim}
The function $\gamma \rightarrow \beta^{'}_{\text{min}}(\gamma)$ is decreasing when $\gamma$ varies from $0$ to $1$. Consequently, 
$\beta^{'}_{\text{min}}(\gamma)$ is strictly inferior to  $1/2$ for all values of $\gamma$ if and only if 
$\beta_{\text{min}}(0) < \frac{1}{2}$. 
\end{proposition}

The expression of  $\beta^{'}_{\text{min}}(0)$ can be deduced directly from the one of $\beta_{\text{min}}(0)$ (\ref{eq:beta-min-zero}) : 
\begin{equation}
\label{eq:beta-min-prim-zero}
\beta^{'}_{\text{min}}(0) = \min_{f_a \in \mathcal{F}([-\frac{1}{2T}, \frac{1}{2T}])} \kappa T  \frac{ \int_{\mathbb{R}} |f_{a}(t)|^4 dt}{(\int_{\mathbb{R}} |f_{a}(t)|^2 dt)^2} + 2 = \eta_{\text{min}} +2
\end{equation}
with $\eta_{\text{min}}$ given by equation \eqref{eq:expre-etamin}.

\begin{figure}[ht]
  \centerline{\includegraphics[width=7cm]{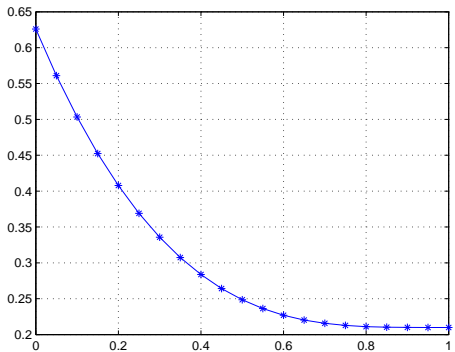}}
\caption{$\beta^{'}_{\text{min}}$ as a function of $\gamma$ for a BPSK signal.}
\label{fig:betamin_modifie_bpsk}
\end{figure}

Recall that we can numerically evaluate the values of $\beta^{'}_{\text{min}}$ and $\eta_{\text{min}}$ for all excess bandwidth factor $\gamma\in [0,1]$. Particularly for an excess bandwidth factor of $0$, $\eta_{\text{min}}\simeq -1.36$ and  $\beta^{'}_{\text{min}}=0.64\geq 1/2$. In order to verify the existence of some  $\beta^{'}_{\text{min}}$ values smaller than $1/2$ we present in figure \ref{fig:betamin_modifie_bpsk} the graph of $\beta^{'}_{\text{min}}(\gamma)$ for all excess bandwidth factor $\gamma\in [0,1]$. The figure confirms the decreasing nature of  $\beta^{'}_{\text{min}}$ with respect to $\gamma$ and shows that $\beta^{'}_{\text{min}} < 1/2$ as soon as $\gamma > 0.1$. Consequently, we are sure to separate the  BPSK sources using the minimization of $J^{'}(r)$ if their common excess bandwidth factor is superior to 0.1. 
  
When the excess bandwidth factor is inferior to 0.1, the inequality $J^{'}(r) \geq m(r)$ does not allow any conclusion to be drawn as to the global minimum of $J^{'}(r)$. However, in such cases, we can consider the approach used in \cite{jal-che-lou-06-soumis} in the case of circular signals
and inequality (\ref{eq:minoration-l'}). After some algebra, we can prove that if $\beta^{'}_{\text{min}} > 1/2$, then the global minimum of $J^{'}(r)$ is reached for filters which allow the extraction of one of the sources, if the following 2 sufficient conditions are met.

\begin{equation}
\label{ameliore:cond1'}
\begin{cases}
&\eta_{\text{min}} + 3 - (K+1) (\beta^{'}_{{\text{min}}} - \frac{1}{2})  >  0\\
&\beta^{'}_{\text{min}} \left(  K\beta^{'}_{\text{min}} - \frac{1}{2} \right) - \\
&(K-1) \left( 2 - \sqrt{\frac{3}{2}} \sqrt{(K(\beta^{'}_{{\text{min}}} -\frac{1}{2}) - (\eta_{\text{min}} + \frac{3}{2}) }\right)^{2} <  0
\end{cases}
\end{equation} 
We give the proof of these conditions in \ref{ap:C}.
We can easily verify that these conditions hold for $\gamma \in [0, 0.1]$ if the number of sources $K$ is inferior to 10, which is very satisfying in the considered context. 

\section{The case of general mixtures}
\label{sec:Rem}

\subsection{Generalisation of $J^{'}(r)$}

The results obtained in the case of a mixture of BPSK signals sharing the same characteristics can be extended to more general mixtures of circular linearly modulated signals and BPSK signals. 
The logic behind the definition of  $J^{'}(r)$ is to subtract from $J(r)$ the square modulus of the non conjugate cyclic correlation coefficients at time lag 0 and at the non conjugate cyclic frequencies  $\{ 2 \delta f_k, s_k \, \mathrm{BPSK}\}$. These frequencies are called in the following the significant non conjugate cyclic frequencies of the received signal, and we denote by  $I_{c,s}$  this set. The definition of $J^{'}(r)$ thus becomes 
\begin{align}
\label{eq:def-J'-general}
J^{'}(r) &=  J(r) - \sum_{\alpha_c \in I_{c,s}} |R^{(\alpha_c)}(0)|^{2}\\
& =  < \mathbb{E} \left(|r(n)|^{2} - 1 \right)^{2} > - \sum_{\alpha_c \in I_{c,s}} \left| < \mathbb{E}(r^{2}(n)) e^{-2 i \pi n \alpha_c}> \right|^{2}  \nonumber
\end{align}
We assume that the mixture contains $L$ groups 
of $(K_l)_{l=1, \ldots, L}$ BPSK signals sharing the same characteristics (symbol period, carrier frequency, excess bandwith) and linearly modulated circular source signals whose symbol 
period differ from those of the BPSK signals. If source $k$ is circular, then it holds that $\beta^{'}_{min,k} = \beta_{min,k} \geq 1 > \beta^{'}_{min}$. Therefore, it is easy to check that $J^{'}(r) \geq m(r)$ where $m(r)$ is still defined by (\ref{eq:def-m}). Proposition \ref{prop:separation-beta<1/2} thus implies that if the excess bandwith of the BPSK signals are greater than 0.1, then the minimization of $J^{'}$ allows to extract all the BPSK signals. The case where some of these excess bandwiths  are less than 0.1 is more difficult, but could be addressed using the previous approach. We just mention that if the cyclic and non conjugate cyclic frequencies of the sources are pairwise different, then the minimization of $J^{'}$ still allows to extract the $K$ sources whatever their excess bandwiths. In effect, $J^{'}(r)$ can be expressed as 
\begin{equation}
\label{eq:Jprim}
J^{'}(r) = \sum_{k=1}^{K} \beta^{'}(\tilde{s}_k) \|f_k\|^{4} + 2 \sum_{k_1 \neq k_2} \|f_{k_1}\|^{2}  \|f_{k_2}\|^{2}
- 2 \sum_{k=1}^{K} \|f_k\|^{2} + 1
\end{equation}

If the source $k$ is circular $\beta^{'}(\tilde{s}_k)$ coincides with $\beta(\tilde{s}_k)$ (\ref{eq:def-beta}) while for a BPSK source $\beta^{'}(\tilde{s}_k)$ is defined by (\ref{eq:def-beta'-BPSK}). The expression of $J^{'}(r)$ is therefore similar to that of $J(r)$, and thus all results obtained in section \ref{sec:differentes} remain valid since for all $k$, $\beta^{'}_{k,min} \leq \beta_{k,min} < 2$. 
The modification of $J$ proposed in order to solve the problems generated by mixtures of non circular sources of the same nature thus does not modify the results obtained in the context of circular or non circular sources having different cyclic and non conjugate cyclic frequencies. \\

\subsection{Frequency offset estimation}
The use of $J^{'}$ requires of course the correct estimation of the significant non conjugate cyclic frequencies of the received signal prior the source separation. Fortunately, this is a much easier task than the estimation of the baud rates, because the non conjugate cyclic correlation coefficients of the received signal at twice the frequency offsets are not affected by possible low excess bandwidths of the source signals. 
A simple detection technique based on the examination of the modulus of the periodogram of the signal $(y_m(n+\tau) y_m(n))_{n \in \mathbb{Z}}$ (see for example \cite{r-cib-lou-ser-gia-02-1}) may be successfully used. We also notice that if the estimation algorithm detects not only the significative non-conjugate cyclic frequencies $\{ 2 \delta f_k, s_k \, \mathrm{BPSK} \}$, but some non significative conjugate cyclic frequencies such as $2 \delta f_{k_0} + \alpha_{k_0}$ or  $2 \delta f_{k_0} - \alpha_{k_0}$, then the behaviour of function $J^{'}$ is even better because $\beta^{'}(\tilde{s}_{k_0})$ defined in principle by (\ref{eq:def-bis-beta'-BPSK}) is replaced by $\beta(\tilde{s}_{k_0}) - |R^{(0)}_{c,\tilde{s}_{k_0}}|^{2} -  |R^{(\alpha_{k_0})}_{c,\tilde{s}_{k_0}}|^{2}$ or $\beta(\tilde{s}_{k_0}) - |R^{(0)}_{c,\tilde{s}_{k_0}}|^{2} -  |R^{(-\alpha_{k_0})}_{c,\tilde{s}_{k_0}}|^{2}$. $\beta^{'}_{min,k_0}$ is thus lower than what is predicated by Figure \ref{fig:betamin_modifie_bpsk}. The sufficient condition $\beta^{'}_{min,k_0} \geq \frac{1}{2}$ is thus less restrictive than in the case where  $2 \delta f_{k_0} + \alpha_{k_0}$ and $2 \delta f_{k_0} - \alpha_{k_0}$ are not detected. 

\section{Simulations}
\label{sec:sim}
\subsection{Implementation of the deflation approach} 

In order to introduce the deflation approach we have implemented, we consider  $\hat{{\bf g}}^{1}$ the extracting filter obtained by minimizing the cost function ($J(r)$ or $J^{'}(r)$). We denote by 
$$\hat{r}_1(m) = \lceil\hat{{\bf g}}^{1}(z)\rceil {\bf y}(m)$$
an estimator of a filtered version of one of the source signals. 
The deflation approach consists in subtracting the contribution of this particular source from the observed signal ${\bf y}(m)$. As a result a new signal ${\bf y}^{(2)}(m)$ is formed containing only the contributions of the other sources. We can then run the extraction algorithm on ${\bf y}^{(2)}(m)$ in order to determine a new filter $\tilde{{\bf g}}^{(2)}(z)$ for which, signal
$$\tilde{r}_2(m) = [\tilde{{\bf g}}^{(2)}(z)] {\bf y}^{(2)}(m)$$
represents an estimator of a filtered version of a second source.
In practice, the first subtraction is not perfect and $\hat{r}_1(m)$ may contain residual filtered versions of the remaining sources. This can render the convolutive mixture defined by ${\bf y}^{2}(m)$ more difficult to inverse than the original one. It is therefore reasonable to try to go back to the originally mixture ${\bf y}(m)$, and apply the extraction algorithm initialised with a filter close enough to ${\bf g}^{(2)}(z)$, the filter that allows the extraction of the second source. This initial filter, denoted by ${\bf g}_{init}^{(2)}(z) = \sum_{l=0}^{L} {\bf g}^{(2)}_{init,l} z^{-l}$, is obtained by minimizing with respect to ${\bf g}_{init}^{(2)} = ({\bf g}^{(2)}_{init,0}, \ldots, {\bf g}^{(2)}_{init,L})$ the quadratic criterion
\[
\frac{1}{M} \sum_{m=0}^{M-1} \left| [{\bf g}^{(2)}_{init}(z)] {\bf y}(m) - \tilde{r}_2(m) \right|^{2} 
\]
This initialization, proposed in \cite{tugnait-1999}, allows the extraction of the second source from the original mixture ${\bf y}$ to be achieved with better performance.


\subsection{Simulations parameters}
The experimental results we present in the following were obtained in the context of blind separation of a convolutive mixture of $K=3$ equal power BPSK  modulated signals, observed by a receiver equipped with a circular array of $N=5$ sensors distanced from one another by half a wavelength. All sources have the same excess bandwidth factor $\gamma=0.5$.

The propagation channels are multi path and affected by a Rayleigh fading. An arbitrary path ($k$) is characterized by its delay $\tau_k$, elevation $\phi_k$, azimuth $\theta_k$ and attenuation $\lambda_k$. We consider the ETSI channels BUx, TUx, HTx, RAx. For each experiment, the arrival angles on the different paths ($\phi$ and $\theta$) of the signals are randomly chosen inside $[-\pi/2, \pi/2]$ and $[-\pi, \pi]$ respectively. The different complex amplitudes on each path are also randomly chosen for each experiment. Generating different channel characteristics from one experiment to another enables us to have statistically significant results. We suppose that the central frequency of the receive filter of the receiver is $f_0 = 1GHz$ and
that the received signal is corrupted by a white, additive complex gaussian noise with power spectral density $N_0$. The signal to noise ratio per source signal $\frac{E_s}{N_0}$ is equal to 20 dB. We have considered two opposite scenarios : 
\begin{itemize}
\item all BPSK signal have the same symbol period $T = 3.6 \mu s$ and same frequency offsets ($\delta f$)
\item the BPSK signal have different symbol periods ($T_1 = 3.4\mu s , T_2=3.6 \mu s , T_3 = 3.9 \mu s$ ) and different frequency offsets ($\delta f_1 \neq \delta f_2 \neq \delta f_3$)
\end{itemize}
In both cases, the sampling period $T_e$ is equal to $\frac{T}{1.6}$, and the carrier frequency offsets are randomly chosen on each trial such that the generated signals satisfy the sampling theorem. We also considered different observation durations $T_{obs}=2000T$, $T_{obs}=1000T$ and $T_{obs}=500T$ for the initial received signal used to compute the separating filters and a longer observation duration of $T_{perf}=20000T$ for the performance analysis. For each possible type of mixture we considered $1000$ independent experiments.

\subsection{Numerical results}

We chose two metrics of performance for our separating method : the signal to interference plus noise ratio (SINR) at the output of the separating filter ${\bf g}$  and the symbol error rate (SER) computed after applying a blind CMA fractional equalizer, supposed to know the baud rates and the carrier frequency offsets of the sources, to the separated signal. In order to compare the different separating algorithms we consider the number of experiments where we obtain a SER inferior to $10^{-2}$.

Moreover, since the channels are randomly selected from one experiment to another, we need a reference measure of the difficulty of the separation problem. We chose to compute, for each source $k$  the performances obtained in a non blind context with the minimum mean square estimator (MMSE). The Wiener filter $\hat{{\bf g}}^{(k)}_{wiener}(z)$ obtained with this method is a finite impulse response filter of the same size as $\hat{{\bf g}}$. This filter is chosen non causal, and its coefficients are estimated from the samples of the received signal $({\bf y}(m))_{m=0, \ldots, M-1}$ and those of the transmitted signal $(s_k(m))_{m=0, \ldots, M-1}$ as if the receiver worked with a learning sequence of $M$ samples. The performances provided by this filter thus represent an upper bound as to what we could achieve in a blind context. \\

\begin{figure}[!h]
         \includegraphics[scale=0.45]{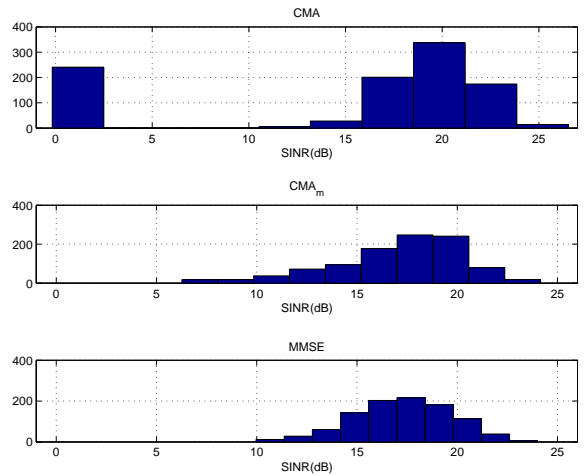}      
  \caption{Histogram of SINR obtained after extracting one source from a mixture of 3 identical BPSK signals, sent over a BUx channal and observed for a duration $T_{obs}=2000T$.}
   \label{fig:3bpsk_sinr_Bux2000}
\end{figure}

Table \ref{tab:SINR_3BPSK_tout_commun} contains the results associated with the first scenario. Notice that the number of times where the SER corresponding to the separation method based on the CMA algorithm is inferior to $10^{-2}$ is smaller than the one corresponding to the separation with the modified CMA criterion. This is due to the large number of cases where the CMA algorithm does not correctly extract the sources from the mixture. Contrariwise, the modified CMA algorithm succeeds in extracting one source from the mixture. This phenomenon is visible in figure \ref{fig:3bpsk_sinr_Bux2000} where we present the histograms of the SINR obtained after the extraction of one source from the mixture using the   CMA, modified CMA and MMSE methods, when considering BUx type communication channels and a duration of observation of $2000T$. It is easy to see that in an important number of cases the SINR values corresponding to the CMA method are close to 0 dB meaning that no source was correctly extracted. The modified CMA algorithm significantly reduces the number of unsuccessful extractions and its performance is close to that of the Wiener filter (MMSE). This phenomenon can also be observed on the results obtained on the other channels and when the duration of observation is smaller. \\

The results obtained in the second scenario are presented in table \ref{tab:SINR_3BPSK_tout_dif}. Notice that in this case the performance of the CMA extraction method is very close to that of the modified CMA method but does not generally surpass it. This confirms the good behaviour of the CMA algorithm when separating mixtures of signals with different characteristics all the while showing that the use of the modified CMA algorithm in such cases could bring some improvement. Generally, the performance of the two blind separation methods are close to that of the MMSE method. 

\begin{table*}[ht]
\caption{Percentage of TES $<10^{-2}$ for a mixture of 3 BPSK signals whose cyclic and non conjugate cyclic frequencies are (a) identical and (b) all different.}
\begin{center}
 \subtable[]{
             \scalebox {0.8}[0.8]{ 
\begin{tabular}{||l||l|l|l||l|l|l||l|l|l||}
\hline 
\textbf{No.symboles} & \multicolumn{3}{c||}{\color{red}{\textbf{2000}}} & \multicolumn{3}{c||}{\color{red}{\textbf{1000}}} & \multicolumn{3}{c||}{\color{red}{\textbf{500}}}\\
\hline 
\color{red}{\textbf{BUx:}} &  \textbf{\small{BPSK}} &  \textbf{\small{BPSK}} &  \textbf{\small{BPSK}} &  \textbf{\small{BPSK}} &  \textbf{\small{BPSK}} &  \textbf{\small{BPSK}} &  \textbf{\small{BPSK}} &  \textbf{\small{BPSK}} &  \textbf{\small{BPSK}}\\
\hline
\textbf{CMA} & 84\% & 83\% & 83.8\%    & 82.6\% & 83.2\% & 81.8\%   & 83.2\% & 81.8\% & 84.7\% \\ 
\hline
\textbf{CMAm} & 99.9\% & 100\% & 100\% & 100\% & 100\% & 99.8\%   & 97.4\% & 95\% & 97.2\% \\ 
\hline
\textbf{MMSE} & 100\% & 99.8\% & 100\% & 99.9\% & 100\% & 99.9\%    & 99.8\% & 100\% & 100\%\\ 
\hline  
\color{red}{\textbf{TUx:}} &  \textbf{\small{BPSK}} &  \textbf{\small{BPSK}} &  \textbf{\small{BPSK}} &  \textbf{\small{BPSK}} &  \textbf{\small{BPSK}} &  \textbf{\small{BPSK}} &  \textbf{\small{BPSK}} &  \textbf{\small{BPSK}} &  \textbf{\small{BPSK}}\\
\hline
\textbf{CMA} & 88.9\% & 87\% & 89.2\%   & 89.4\% & 85.2\% & 86.1\%  & 86.8\% & 86.8\% & 87.6\% \\ 
\hline
\textbf{CMAm} & 99.9\% & 100\% & 100\% & 99.8\% & 99.8\% & 99.6\%  & 95.7\% & 95.3\% & 94.5\% \\ 
\hline
\textbf{MMSE} & 100\% & 99.8\% & 100\%  & 100\% & 100\% & 99.9\%  & 100\% & 100\% & 100\%  \\ 
\hline
\color{red}{\textbf{HTx:}} &  \textbf{\small{BPSK}} &  \textbf{\small{BPSK}} &  \textbf{\small{BPSK}} &  \textbf{\small{BPSK}} &  \textbf{\small{BPSK}} &  \textbf{\small{BPSK}} &  \textbf{\small{BPSK}} &  \textbf{\small{BPSK}} &  \textbf{\small{BPSK}}\\
\hline
\textbf{CMA} & 89\% & 87.2\% & 86\% & 87.6\% & 88\% & 88.4\%   & 87.5\% & 87.7\% & 86.1\% \\ 
\hline
\textbf{CMAm} & 99.7\% & 99.8\% & 100\%  & 99.2\% & 99.5\% & 99.4\% & 91.5\% & 91.5\% & 91.4\%\\ 
\hline
\textbf{MMSE} & 99.9\% & 100\% & 100\% & 100\% & 100\% & 100\% & 100\% & 100\% & 100\%\\ 
\hline
\color{red}{\textbf{RAx:}} &  \textbf{\small{BPSK}} &  \textbf{\small{BPSK}} &  \textbf{\small{BPSK}} &  \textbf{\small{BPSK}} &  \textbf{\small{BPSK}} &  \textbf{\small{BPSK}} &  \textbf{\small{BPSK}} &  \textbf{\small{BPSK}} &  \textbf{\small{BPSK}}\\
\hline
\textbf{CMA} & 78\% & 79.7\% & 79\%   & 78.9\% & 81.3\% & 79.6\% & 81\% & 81.1\% & 80.3\%\\ 
\hline
\textbf{CMAm} & 100\% & 99.9\% & 99.9\%  & 99.4\% & 99.1\% & 99.2\%  & 93.6\% & 94.9\% & 92.9\%\\ 
\hline
\textbf{MMSE} & 100\% & 99.9\% & 99.9\%  & 100\% & 100\% & 100\% & 100\% & 100\% & 99.9\% \\ 
\hline
\end{tabular} 
\label{tab:SINR_3BPSK_tout_commun}
         }  
 }
 \subtable[ ]{
             \scalebox {0.8}[0.8]{ 
             \begin{tabular}{||l||l|l|l||l|l|l||l|l|l||}
             \hline 
\textbf{No.symboles} & \multicolumn{3}{c||}{\color{red}{\textbf{2000}}} & \multicolumn{3}{c||}{\color{red}{\textbf{1000}}} & \multicolumn{3}{c||}{\color{red}{\textbf{500}}}\\
\hline 
\color{red}{\textbf{BUx:}} &  \textbf{\small{BPSK}} &  \textbf{\small{BPSK}} &  \textbf{\small{BPSK}} &  \textbf{\small{BPSK}} &  \textbf{\small{BPSK}} &  \textbf{\small{BPSK}} &  \textbf{\small{BPSK}} &  \textbf{\small{BPSK}} &  \textbf{\small{BPSK}}\\
\hline
\textbf{CMA}    & 100\% & 99.9\% & 99.8\%   & 99.8\% & 99.7\% & 99.8\%   & 99.5\% & 99.4\% & 99.6\% \\ 
\hline
\textbf{CMAm}   & 100\% & 99.9\% & 100\%    & 100\% & 99.9\% & 99.9\%   & 99.7\% & 99.5\% & 99.8\%    \\ 
\hline
\textbf{MMSE}   & 100\% & 100\% & 100\%     & 100\% & 100\% & 100\%    & 100\% & 100\% & 100\%   \\ 
\hline
\color{red}{\textbf{TUx:}} &  \textbf{\small{BPSK}} &  \textbf{\small{BPSK}} &  \textbf{\small{BPSK}} &  \textbf{\small{BPSK}} &  \textbf{\small{BPSK}} &  \textbf{\small{BPSK}} &  \textbf{\small{BPSK}} &  \textbf{\small{BPSK}} &  \textbf{\small{BPSK}}\\
\hline
\textbf{CMA}  & 99.6\% & 99.5\% & 99.7\%   & 99.3\% & 99.5\% & 99.6\%  & 99.7\% & 99.4\% & 99.2\%    \\ 
\hline
\textbf{CMAm} & 100\% & 99.9\% & 99.7\%    & 99.7\% & 99.7\% & 99.6\%  & 98.1\% & 98.9\% & 98.4\%    \\ 
\hline
\textbf{MMSE} & 100\% & 99.9\% & 100\%     & 100\% & 100\% & 100\%     & 100\% & 100\% & 100\% \\ 
\hline
\color{red}{\textbf{HTx:}} &  \textbf{\small{BPSK}} &  \textbf{\small{BPSK}} &  \textbf{\small{BPSK}} &  \textbf{\small{BPSK}} &  \textbf{\small{BPSK}} &  \textbf{\small{BPSK}} &  \textbf{\small{BPSK}} &  \textbf{\small{BPSK}} &  \textbf{\small{BPSK}}\\
\hline
\textbf{CMA}  & 98.8\% & 98.7\% & 98.8\%   & 98\% & 97.8\% & 98.1\%    & 96.7\% & 96\% & 96\%    \\ 
\hline
\textbf{CMAm} & 100\%  & 99.8\% & 100\%    & 99.5\% & 99.4\% & 99.4\%  & 98\%   & 98.2\% & 97.7\%   \\ 
\hline
\textbf{MMSE} & 100\% & 100\%   & 100\%    & 100\% & 100\% & 100\%     & 100\%  & 100\% & 100\% \\ 
\hline
\color{red}{\textbf{RAx:}} &  \textbf{\small{BPSK}} &  \textbf{\small{BPSK}} &  \textbf{\small{BPSK}} &  \textbf{\small{BPSK}} &  \textbf{\small{BPSK}} &  \textbf{\small{BPSK}} &  \textbf{\small{BPSK}} &  \textbf{\small{BPSK}} &  \textbf{\small{BPSK}}\\
\hline
\textbf{CMA}    & 98.3\% & 98.9\% & 98.3\%    & 99.2\% & 98.9\% & 98.5\%  & 99\% & 98.8\% & 98.7\%  \\ 
\hline
\textbf{CMAm}   & 99.8\% & 99.8\% & 100\%     & 98.9\% & 99\% & 99\%      & 98.4\% & 98.2\% & 98.3\%   \\ 
\hline
\textbf{MMSE}   & 99.9\% & 100\%  & 100\%     & 100\% & 100\% & 100\%     & 100\% & 100\% & 100\%  \\ 
\hline
\end{tabular}  
        \label{tab:SINR_3BPSK_tout_dif}
                          }  
             }               
   \end{center}
\end{table*}
$$ $$

\section{Conclusion}

We investigated the separation of convolutive mixtures of second order circular linearly modulated signals and BPSK signals in the context of passive listening. We considered only deflation approaches coupled with the minimization of the CMA cost function. We proved that if the different source signals do not share the same cyclic and non conjugate cyclic frequencies, the minimization of the CMA cost function ensures the extraction of a filtered version of one of the source signals. We have also shown that in this case and under a condition which is always verified in practice, all the local minima of the CMA criterion are separating points. This result is no longer true when mixtures of BPSK signals sharing the same baud rate and carrier frequency are considered.  In this case we have shown the existence of non separating  local minima of the CMA cost function that prove to be quite attractive. A modification of the CMA criterion was proposed, based on the knowledge of the most significant non conjugate cyclic frequencies of the received signal. Moreover, the minimization of this new criterion was also proved to be a reliable approach in a much more general context. 

\appendix
\section{Proof of Proposition \ref{prop:betaminBPSK} }
\label{ap:A}
Although the proof of this proposition is very similar to the one in \cite{jal-che-lou-06-soumis}, we provide it in order to make the paper reasonably self-contained. For simplicity reasons,  we assume the carrier frequency offsets to be 0. This assumption does not reduces the generality of the results.
\begin{proposition}
\label{prop:f-num-analog}
Suppose that $T_e$ is not a multiple of $T/2$ (this automatically holds from the hypothesis) and that $T_e<T/(1+\gamma)$ (this holds since(\ref{eq:Shannon}) holds). Let $f(z)$ be a transfer function for which $\|f\|\leq\infty$. If $g_a(t)$ denotes the shaping filter of signal $s_a(t)$, then the function $\hat{f}_{a}(\nu)$ defined as
\begin{equation} 
\label{eq:def-fa}
\hat{f}_{a}(\nu)=f_{}( e^{2i\pi \nu T_{e}} )\hat{g}_{a}(\nu),\,\forall \nu\in \mathbb{R}
\end{equation}
vanishes outside $\mathcal{B}=\left[-\frac{1+\gamma}{2T},\frac{1+\gamma}{2T} \right] $ and belongs to the space $\mathcal{F}(\mathcal{B})$.
Let $f_{a}(t)$ be its inverse Fourier transform in the $\mathbb{L}^2$--sense. For every, $t$ we define the continuous-time signal $(r_{a}(t))_{t\in\mathbb{R}}$ as
$$ r_{a}(t)=\sum_{j\in\mathbb{Z}}{a(j)f_{a}(t-jT)} $$  
Then, the discrete-time signal $r(n)=[f(z)]s(n)$ coincides with the discrete-time signal $r_{a}(nTe)$.
\end{proposition}

This result is proved in  \cite{hou-che-lou} when the filter $f(z)$ has a summable impulse response and in \cite{jal-che-lou-06-soumis} when 
$\|f\|\leq\infty$.
Using proposition \ref{prop:f-num-analog}, we have
$\forall f(z), \, \exists f_{a}(t)$ such that  $<\EE\left|r(n)\right|^{2}> = <\EE\left|r_{a}(nT_{e})\right|^{2}>$ 
and
$< \EE\left\{ r(n)^{2}\right\}> =< \EE\left\{ r_{a}(nT_{e})^{2}\right\}> $.
Considering the time average of the Fourier series expansion of $\EE\left|r_{a}(t)\right|^{2}$ and $\EE\left\{ r_{a}(t)^{2}\right\} $, when $T\not\in \left\{ T,\frac{T}{2},\frac{T}{3},\frac{2T}{3}  \right\}$ we get

\begin{align}
&< \EE\left|r(n)\right|^{2}>  = < \EE\left|r_{a}(nT_{e})\right|^{2}> \nonumber\\
&=R_{r_{a}}^{(0)}(0)+R_{r_{a}}^{(\frac{1}{T})}(0)< e^{2i\pi\frac{nT_e}{T}}> +R_{r_{a}}^{(\frac{1}{T})}(0)< e^{-2i\pi\frac{nT_e}{T}}> \nonumber\\
&=R_{r_{a}}^{(0)}(0)=\frac{1}{T}\int_{\mathbb{R}}\left|f_{a}(t)\right|^{2}dt\label{pet}
\end{align}

\begin{align}
&< \EE\left\{ r(n)^{2}\right\} >  =  < \EE\left\{ r_{a}(nT_{e})^{2}\right\} > \nonumber\\
&=  R_{c,r_{a}}^{(0)}(0)+R_{c,r_{a}}^{(\frac{1}{T})}(0)< e^{2i\pi\frac{nT_e}{T}}> +R_{c,r_{a}}^{(-\frac{1}{T})}(0)< e^{-2i\pi\frac{nTe}{T}}> \nonumber\\
&=R_{c,r_{a}}^{(0)}(0)= \frac{1}{T}\int_{\mathbb{R}}f_{a}(t)^{2}dt\label{pet2}
\end{align}

Furthermore

\begin{align}
&R_{r_{a}}^{(\pm\alpha)}(0)=\frac{1}{T}\int_0^T{\EE\left|r_{a}(nT_{e})\right|^{2}e^{\mp 2i\pi\frac{t}{T}} dt} = \frac{1}{T}\int_{\mathbb{R}}\left|f_{a}(t)\right|^{2}e^{\mp2i\pi\frac{t}{T}}dt\nonumber\\
&R_{c,r_{a}}^{(\pm\alpha_c)}(0)=\frac{1}{T}\int_0^T{\EE\left|r_{a}(nT_{e})\right|^{2}e^{\mp2i\pi\frac{t}{T}} dt}=\frac{1}{T}\int_{\mathbb{R}}f_{a}(t)^{2}e^{\mp2i\pi\frac{t}{T}}dt 
\label{eq:multiform}
\end{align}

A similar reasoning can be carried on for the 4-th order cumulant of signal $r_a$: function $c_{4}(r_{a}(t))$ can be written as $c_{4}(r_{a}(t))=\kappa \sum_{n}\left|f(t-nT\right|^{4}$ where $\kappa$ is the 4-th order cumulant of the transmitted symbol sequence. This function is periodic of period $T$ and, due to the limited bandwidth of filter $f_{a}$, has at most 7 cyclic frequencies. Its Fourier series expansion therefore is :
$$c_{4}(r_{a}(t))=\sum_{k=-3}^{3}c_{k}e^{2i\pi k\frac{t}{T}}$$
Because of the conditions imposed on $T_{e}$, the terms \mbox{$< e^{2i\pi k\frac{nT_{e}}{T}}> $} are zero if  $k\neq0$, which means that:

\begin{equation}
< c_{4}(r(n))> =c_{0}=\frac{\kappa}{T}\int_{\mathbb{R}}\left|f_{a}(t)\right|^{4}dt
\label{pet3}
\end{equation}
where $c_{0}$ is the constant value of the Fourier series expansion of $< c_{4}(r_{a}(t))> $.

Applying (\ref{pet}), (\ref{pet2}), (\ref{eq:multiform}) and (\ref{pet3}) to signal \mbox{$\tilde{s} = \left\lceil \frac{f(z)}{\|f\|} \right\rceil s(n)$} and recalling that
\begin{align}
\| f \|^ 2 &= \int_{-1/2}^{1/2}{|f(e^{2i\pi\nu})|^2S_{s}^{(0)}}d\nu = <\EE|\lceil f(z)\rceil s(n)|^2>\nonumber\\
 &= <\EE|\lceil f(z)\rceil s_{a}(nTe)|^2>=\frac{1}{T}\int_{\mathbb{R}}{\left|f_{a}(t)\right|^{2}dt}
\end{align}
we evaluate the terms in expression (\ref{eq:expre-betamin}) and find the expression of  $\Phi(f_a)$.
\section{Proof of  Proposition \ref{prop:minoration-l'} }
\label{ap:B}
In the following, we consider once more filter $\hat{f}_a(\nu)$ defined in \eqref{eq:def-fa} 
We begin by expressing the cyclic and non conjugate cyclic correlation coefficients involved in expression (\ref{eq:minoration-l'}) as 
\begin{align}
&R_{\tilde{s}}^{(\alpha)}(0)=\frac{\int_{-\frac{1+\gamma}{2T}}^{\frac{1+\gamma}{2T}}\hat{f}_{a}(\nu)\hat{f}_{a}(\nu-\frac{1}{T})^{*}d\nu}{\int_{-\frac{1+\gamma}{2T}}^{\frac{1+\gamma}{2T}}\left|\hat{f}_{a}(\nu)\right|^{2}d\nu} \nonumber\\
&R_{c,\tilde{s}}^{(+\alpha)}(0)=\frac{\int_{-\frac{1+\gamma}{2T}}^{\frac{1+\gamma}{2T}}\hat{f}_{a}(\nu)\hat{f}_{a}(\frac{1}{T}-\nu)d\nu}{\int_{-\frac{1+\gamma}{2T}}^{\frac{1+\gamma}{2T}}\left|\hat{f}_{a}(\nu)\right|^{2}d\nu}\nonumber\\
&R_{c,\tilde{s}}^{(-\alpha)}(0)=\frac{\int_{-\frac{1+\gamma}{2T}}^{\frac{1+\gamma}{2T}}\hat{f}_{a}(\nu)\hat{f}_{a}(-\nu-\frac{1}{T})d\nu}{\int_{-\frac{1+\gamma}{2T}}^{\frac{1+\gamma}{2T}}\left|\hat{f}_{a}(\nu)\right|^{2}d\nu}\label{et.mul1}
\end{align}
In order to simplify the notations, we denote by  $B_{\gamma}^{+}$ and respectively $B_{\gamma}^{-}$ the intervals $B_{\gamma}^{+}=\left[\frac{1-\gamma}{2T},\frac{1+\gamma}{2T}\right]$ and $B_{\gamma}^{-}=\left[-\frac{1+\gamma}{2T},-\frac{1-\gamma}{2T}\right]$. It is straightforward that
\begin{equation}
\int_{-\frac{1+\gamma}{2T}}^{\frac{1+\gamma}{2T}}\left|\hat{f}_{a}(\nu)\right|^{2}d\nu\geq\int_{B_{\gamma}^{-}}\left|\hat{f}_{a}(\nu)\right|^{2}d\nu+\int_{B_{\gamma}^{+}}\left|\hat{f}_{a}(\nu)\right|^{2}d\nu\label{b.11}\end{equation}
Notice that the functions $\nu\rightarrow\hat{f_{a}}(\nu)\hat{f_{a}}(\nu-\frac{1}{T})^{*}$
and $\nu\rightarrow\hat{f_{a}}(\nu)\hat{f_{a}}(\frac{1}{T}-\nu)$ are zero unless $\nu\in B_{\gamma}^{+}$ and that function $\nu\rightarrow\hat{f_{a}}(\nu)\hat{f_{a}}(-\nu-\frac{1}{T})$ is also zero outside of $B_{\gamma}^{-}$. This means that 
\begin{align}
&\int_{-\frac{1+\gamma}{2T}}^{\frac{1+\gamma}{2T}}\hat{f_{a}}(\nu)\hat{f_{a}}(\nu-\frac{1}{T})^{*}d\nu=\int_{B_{\gamma}^{+}}\hat{f_{a}}(\nu)\hat{f_{a}}(\nu-\frac{1}{T})^{*}d\nu\nonumber\\
&\int_{-\frac{1+\gamma}{2T}}^{\frac{1+\gamma}{2T}}\hat{f_{a}}(\nu)\hat{f_{a}}(\frac{1}{T}-\nu)d\nu=\int_{B_{\gamma}^{+}}\hat{f_{a}}(\nu)\hat{f_{a}}(\frac{1}{T}-\nu)d\nu\nonumber\\
&\int_{-\frac{1+\gamma}{2T}}^{\frac{1+\gamma}{2T}}\hat{f_{a}}(\nu)\hat{f_{a}}(-\nu-\frac{1}{T})d\nu=\int_{B_{\gamma}^{-}}\hat{f_{a}}(\nu)\hat{f_{a}}(-\nu-\frac{1}{T})d\nu \label{eq:betas}
\end{align}
Using the inequality of Schwartz we immediately obtain 
{\footnotesize{
\begin{align}
\left|\int_{B_{\gamma}^{+}}\hat{f_{a}}(\nu)\hat{f_{a}}(\nu-\frac{1}{T})^{*}d\nu\right|&\leq\left(\int_{B_{\gamma}^{+}}\left|\hat{f_{a}}(\nu)\right|^{2}d\nu\right)^{1/2}\left(\int_{B_{\gamma}^{+}}\left|\hat{f_{a}}(\nu-\frac{1}{T})\right|^{2}d\nu\right)^{1/2}\nonumber\\
&=\left(\int_{B_{\gamma}^{+}}\left|\hat{f_{a}}(\nu)\right|^{2}d\nu\right)^{1/2}\left(\int_{B_{\gamma}^{-}}\left|\hat{f_{a}}(\nu)\right|^{2}d\nu\right)^{1/2}
\end{align}
\begin{align}
\left|\int_{B_{\gamma}^{+}}\hat{f_{a}}(\nu)\hat{f_{a}}(\frac{1}{T}-\nu)d\nu\right|&\leq\left(\int_{B_{\gamma}^{+}}\left|\hat{f_{a}}(\nu)\right|^{2}d\nu\right)^{1/2}\left(\int_{B_{\gamma}^{+}}\left|\hat{f_{a}}(\frac{1}{T}-\nu)\right|^{2}d\nu\right)^{1/2}\nonumber\\
&=\left(\int_{B_{\gamma}^{+}}\left|\hat{f_{a}}(\nu)\right|^{2}d\nu\right)^{1/2}\left(\int_{B_{\gamma}^{+}}\left|\hat{f_{a}}(\nu)\right|^{2}d\nu\right)^{1/2}\nonumber\\
&=\int_{B_{\gamma}^{+}}\left|\hat{f_{a}}(\nu)\right|^{2}d\nu\label{et.mul2}
\end{align}
\begin{align}
\left|\int_{B_{\gamma}^{-}}\hat{f_{a}}(\nu)\hat{f_{a}}(-\nu-\frac{1}{T})d\nu\right|&\leq\left(\int_{B_{\gamma}^{-}}\left|\hat{f_{a}}(\nu)\right|^{2}d\nu\right)^{1/2}\left(\int_{B_{\gamma}^{-}}\left|\hat{f_{a}}(-\nu-\frac{1}{T})\right|^{2}d\nu\right)^{1/2}\nonumber\\
&=\left(\int_{B_{\gamma}^{-}}\left|\hat{f_{a}}(\nu)\right|^{2}d\nu\right)^{1/2}\left(\int_{B_{\gamma}^{-}}\left|\hat{f_{a}}(\nu)\right|^{2}d\nu\right)^{1/2}\nonumber\\
&=\int_{B_{\gamma}^{-}}\left|\hat{f_{a}}(\nu)\right|^{2}d\nu
\end{align}
}}

Using (\ref{et.mul1}), (\ref{b.11}), (\ref{et.mul2}) and the means inequality ($ab\leq\frac{a^2+b^2}{2}$), we get :
\begin{equation}
|\hat{R}_{\tilde{s}}^{(\alpha)}(0)|\leq 
\frac{\frac{1}{2}\left(\int_{B_{\gamma}^{+}}\left|\hat{f_{a}}(\nu)\right|^{2}d\nu+\int_{B_{\gamma}^{-}}\left|\hat{f_{a}}(\nu)\right|^{2}d\nu\right)}
{\int_{B_{\gamma}^{+}}\left|\hat{f_{a}}(\nu)\right|^{2}d\nu+\int_{B_{\gamma}^{-}}\left|\hat{f_{a}}(\nu)\right|^{2}d\nu}=\frac{1}{2}
\label{eq:inf_demie}
\end{equation}
Since $\hat{R}_{\tilde{s}}^{(-\alpha)}(0)=(\hat{R}_{\tilde{s}}^{(\alpha)}(0))^{*}$  we also get $|\hat{R}_{\tilde{s}}^{(-\alpha)}(0)|\leq\frac{1}{2}$. 

We define the following quantities
\begin{align*}
&x_{i}=\int_{B_{\gamma}^{+}}\left|\hat{f}_{a,k_{i}}(\nu)\right|^{2}d\nu \; ,\; y_{i}=\int_{B_{\gamma}^{-}}\left|\hat{f}_{a,k_{i}}(\nu)\right|^{2}d\nu \, , i\in\{1,2\}\nonumber\\
&t=2\left(R_{\tilde{s}_{k_{1}}}^{(\alpha)}(0)R_{\tilde{s}_{k_{2}}}^{(\alpha)}(0)^{*}+R_{\tilde{s}_{k_{1}}}^{(\alpha)}(0)R_{\tilde{s}_{k_{2}}}^{(\alpha)}(0)^{*}\right)\nonumber\\
&+R_{\tilde{s}_{c,k_{1}}}^{(\alpha)}(0)R_{\tilde{s}_{c,k_{2}}}^{(\alpha)}(0)^{*}+R_{\tilde{s}_{c,k_{1}}}^{(-\alpha)}(0)R_{\tilde{s}_{c,k_{2}}}^{(-\alpha)}(0)^{*}\nonumber
\end{align*}
where $k_1,k_2\in\left\{ 1,2\right\},\;k_1\neq k_2$.

With this notations and using (\ref{et.mul1}),(\ref{b.11}) and (\ref{et.mul2}) for signals $\tilde{s}_{k_1}$ et $\tilde{s}_{k_2}$, we get:
\begin{align}
&|t|\leq4\frac{\sqrt{x_{1}y_{1}}}{x_{1}+y_{1}}\frac{\sqrt{x_{2}y_{2}}}{x_{2}+y_{2}}+\frac{x_{1}}{x_{1}+y_{1}}\frac{x_{2}}{x_{2}+y_{2}}+\frac{y_{1}}{x_{1}+y_{1}}\frac{y_{2}}{x_{2}+y_{2}}\nonumber\\
&\nonumber\\
&|t|\leq\frac{(\sqrt{x_{1}x_{2}}+\sqrt{y_{1}y_{2}})^{2}}{(x_{1}+y_{1})(x_{2}+y_{2})}+\frac{2\sqrt{x_{1}y_{1}}\sqrt{x_{2}y_{2}}}{(x_{1}+y_{1})(x_{2}+y_{2})}\nonumber\\
\end{align}

Using Schwartz inequality
\begin{align}
\sqrt{x_{1}x_{2}}+\sqrt{y_{1}y_{2}}&\leq\sqrt{(\sqrt{x_{1}})^{2}+(\sqrt{y_{1}})^{2}}\sqrt{(\sqrt{x_{2}})^{2}+(\sqrt{y_{2}})^{2}}\nonumber\\
&=\sqrt{(x_{1}+y_{1})(x_{2}+y_{2})}
\end{align}
and since $\frac{\sqrt{x_{i}y_{i}}}{x_{i}+y_{i}}\leq\frac{1}{2}$, we get that$|t| \leq \frac{3}{2}$. 

\section{Proof of conditions \eqref{ameliore:cond1'} }
\label{ap:C}

We pose $\lambda(\tilde{s_{k}})=\sqrt{4\left|R_{\tilde{s}_{k}}^{(\alpha)}(0)\right|^{2}+\left|R_{\tilde{s}_{k}}^{c,(\alpha)}(0)\right|^{2}+\left|R_{\tilde{s}_{k}}^{c,(-\alpha)}(0)\right|^{2}}$. From (\ref{eq:minoration-l'}), we easily get $\lambda(\tilde{s_{k}})\leq\sqrt{\frac{3}{2}}$

Considering the expression (\ref{eq:expre-l'}) of $l^{'}(\tilde{s}_{k_1},\tilde{s}_{k_2})$ it is easy to prove (using Schwartz inequality) that
\begin{equation}
l^{'}(\tilde{s}_{k_{1}},\tilde{s}_{k_{2}})\geq2-\lambda(\tilde{s}_{k_{1}})\lambda(\tilde{s}_{k_{2}})
\label{dem}
\end{equation}

Function $J^{'}(r)$ given by \eqref{eq:expre-J'-developpee} is therefore lower bounded by :

\begin{eqnarray*}
J^{'}(r) & \geq & \sum_{k=1}^{K}\left\Vert f_{k}\right\Vert ^{4}\beta^{'}(\tilde{s}_{k})+\sum_{k_{1}\neq k_{2}}^{K}\left\Vert f_{k_{1}}\right\Vert ^{2}\left\Vert f_{k_{2}}\right\Vert ^{2}(2-\lambda(\tilde{s}_{k_{1}})\lambda(\tilde{s}_{k_{2}})) \\
 &  &  -2\sum_{k=1}^{K}\left\Vert f_{k}\right\Vert ^{2}+1
\label{hi2}
\end{eqnarray*}
 
A better lower bound for $J^{'}(r)$ can be found by choosing a better lower bound of $l^{'}(\tilde{s}_{k_1},\tilde{s}_{k_2})$ and thus a good upper bound for $\lambda(\tilde{s}_{k_{1}})\lambda(\tilde{s}_{k_{2}})$. We first state the following obvious result. 
\begin{lemma}
Let $s_{a,k}(t)=\sum_{l}a_{l,k}g_{a,k}(t-lT_{k})$
be one of the source signals and let $s_{k}(n)$ be the discrete time signal obtained by sampling $s_{a,k}(t)$ at a rate of $T_{e}$. Consider a unit norm filter  $\tilde{f}_{k}(z)$ and signal $\tilde{s}_{k}(n)=\left\lceil \tilde{f}_{k}(z)\right\rceil s_{k}(n)$.
We consider an element $\lambda_{*}\in(0;\sqrt{\frac{3}{2}})$. If  $\lambda(\tilde{s}_{k})\geq\lambda_{*}$, then:
\begin{align*}
\beta^{'}(\tilde{s}_{k}) & \geq\beta^{'}_{*}\end{align*}
with 
\begin{equation}
\label{eq:def_lam}
\beta^{'}_{*}=\eta_{\text{min}}+2+\lambda_{*}^{2}
\end{equation}
where, $\beta^{'}(\tilde{s}_{k})$ and $\eta_{\text{min}}$ are given by (\ref{eq:expre-betamin}) and (\ref{eq:expre-etamin}) respectively.
\end{lemma}

In the following we denote by $\mathcal{K}_1$ the set of index $k$ for which $\lambda(\tilde{s}_k)\leq \lambda_{*}$ and by
$\mathcal{K}_2$ the set of index $k$ for which $\lambda(\tilde{s}_k)> \lambda_{*}$. It is easy to see that this two sets are disjoint and that there union is the set of all source index $\mathcal{K}=\{1\ldots K\}$.
 
Observe that if $k_1, k_2\in \mathcal{K}_1$, then from (\ref{dem})
$l^{'}(\tilde{s}_{k_{1}},\tilde{s}_{k_{2}})\geq 2-\lambda_{*}^2$. If  $k_1, k_2\in \mathcal{K}_2$, then $l^{'}(\tilde{s}_{k_{1}},\tilde{s}_{k_{2}})\geq \frac{1}{2}$, and if $k_1\in\mathcal{K}_1$ and $k_2\in\mathcal{K}_2$, then  $l^{'}(\tilde{s}_{k_{1}},\tilde{s}_{k_{2}})\geq 2 - \sqrt{\frac{3}{2}}\lambda_{*}$.

Under this observations we can give the expression of a better lower bound of $J^{'}(r)$ :
\begin{align}
\label{eq:minor-general}
m(r)=&\beta^{'}_{\text{min}}\sum_{k\in\mathcal{K}_1}{\|f_k\|^4}+(2-\lambda_{*}^2)\sum_{k_1\neq k_2,k_1,k_2\in\mathcal{K}_1}{\|f_{k_1}\|^2\|f_{k_2}\|^2}\nonumber\\
&+\beta_{*}\sum_{k\in\mathcal{K}_2}{\|f_k\|^4}+\frac{1}{2}\sum_{k_1\neq k_2,k_1,k_2\in\mathcal{K}_2}{\|f_{k_1}\|^2\|f_{k_2}\|^2}\nonumber\\
&+2(2-\sqrt{\frac{3}{2}}\lambda_{*})\sum_{k_1\in\mathcal{K}_1,k_2\in\mathcal{K}_2}{\|f_{k_1}\|^2\|f_{k_2}\|^2}\nonumber\\
&-2\sum_{k_1\in\mathcal{K}_1}{\|f_k\|^2}-2\sum_{k_2\in\mathcal{K}_2}{\|f_k\|^2} +1
\end{align}
We consider a vector  $(\| f_k \|)_{k=1, \ldots, K}$ at which the gradient of $m(r)$ is zero. Then, it is easily shown that
\begin{itemize}
\item For each $k\in \mathcal{K}_1$, $\|f_k\|^2$ is either equal to $0$ or to a fixed value (independent of $k$) denoted by $t_1$. We denote by $P_1$ the number of non-zero terms.
\item For each $k\in \mathcal{K}_2$, $\|f_k\|^2$ is either equal to $0$ or to a fixed value (independent of $k$) denoted by $t_2$. We denote by $P_2$ the number of non-zero terms.  
\end{itemize}
If $P_2=0$ then $t_1$ is given by $ t_1=\frac{1}{\beta^{'}_{\text{min}}+(2-\lambda_{*}^2)(P_1-1)}$ and it can be shown that the corresponding value of $m(r)$ increases when $P_1$ increases if \begin{equation}
\label{eq:cond-first}
\beta^{'}_{\text{min}}\leq (2-\lambda_{*}^2)
\end{equation}
If $P_1=0$ then $t_2$ is given by $ t_1=\frac{1}{\beta^{'}_{*}+\frac{1}{2}(P_2-1)}$ and it can be shown that the corresponding value of $m(r)$ decreases when $P_2$ increases.

If $P_1\geq 1$ and $P_2\geq 1$, then $t_1$ and $t_2$ are given by 
{\footnotesize{
\begin{align}
&t_1=\frac{(2-\sqrt{\frac{3}{2}}\lambda_{*})P_2-(\beta_{*}+\frac{1}{2}(P_2-1))}
{(2-\sqrt{\frac{3}{2}}\lambda_{*})^2P_1P_2-(\beta_{\text{min}}+(2-\lambda_{*}^2)(P_1-1))(\beta_{*}+\frac{1}{2}(P_2-1))}\nonumber\\
&t_2=\frac{(2-\sqrt{\frac{3}{2}}\lambda_{*})P_1-(\beta_{\text{min}}+(2-\lambda_{*}^2)(P_1-1))}
{(2-\sqrt{\frac{3}{2}}\lambda_{*})^2P_1P_2-(\beta_{\text{min}}+(2-\lambda_{*}^2)(P_1-1))(\beta_{*}+\frac{1}{2}(P_2-1))}\nonumber\\
\end{align}
}}
The corresponding value of $m(r)$ is
{\footnotesize{
\begin{align}
m=1-\frac{2(2-\sqrt{\frac{3}{2}}\lambda_{*})P_1P_2-(P_1(\beta_{*}+\frac{1}{2}(P_2-1))+P_2(\beta_{\text{min}}+(2-\lambda_{*}^2)(P_1-1) ))}
{(2-\sqrt{\frac{3}{2}}\lambda_{*})^2P_1P_2-(\beta_{\text{min}}+(2-\lambda_{*}^2)(P_1-1))(\beta_{*}+\frac{1}{2}(P_2-1))}
\end{align}
}}and it can be shown that it decreases if $P_2$ increases. Moreover, if condition \eqref{eq:cond-first} holds, then the value of $m$ increases when $P_1$ increases.

The minimum of $m(r)$ therefore corresponds to partitions $(\mathcal{K}_1, \mathcal{K}_2)$ for which $(P_1,P_2)$ are equal to the following three possible values : $(P_1,P_2)=(1,0)$, $(P_1,P_2)=(0,K)$ or $(P_1,P_2)=(1,K-1)$.

It is clear that $m$ coincides with $1-\frac{1}{\beta^{'}_{\text{min}}}$ if $(P_1, P_2)=(1,0)$. Therefore, by Proposition 
\ref{prop:minoration}, separation of a source signal will be achieved if 
\begin{equation}
\label{cond-beta}
m > 1-\frac{1}{\beta^{'}_{\text{min}}}, \; \mbox{for} \;  (P_1,P_2) \neq (1,0)
\end{equation}
In the following, we derive sufficient conditions for which (\ref{cond-beta}) holds. We first consider the case $(P_1,P_2)=(0,K)$, and obtain the following condition 
\begin{equation}
\label{eq:cond-intermerdiaries}
\beta^{'}_{\text{min}}<\frac{\beta_{*}+\frac{1}{2}(K-1)}{K}
\end{equation}
When $(P_1,P_2)=(1,K-1)$ condition \eqref{cond-beta} becomes 
{
\begin{equation}
\label{eq:last}
\frac{(2-\sqrt{\frac{1}{2}})^2(K-1)}{\beta_{*}+\lambda_{*}\frac{1}{2}(K-2)}\geq\beta^{'}_{\text{min}}
\end{equation}
}
Replacing $\lambda_{*}$ from (\ref{eq:def_lam}) in \eqref{eq:cond-first}, \eqref{eq:cond-intermerdiaries} and \eqref{eq:last} we obtain
{
\begin{equation}
\label{eq:cond-fin}
\begin{cases}
&K(\beta^{'}_{\text{min}}-\frac{1}{2})+\frac{1}{2}\leq \beta_{*}\leq 4+\eta_{\text{min}}-\beta^{'}_{\text{min}}\\
&\beta^{'}_{\text{min}}(\beta_{*}+\frac{1}{2}(K-2))-(K-1)\left(2-\sqrt{\frac{3}{2}(\beta_{*}-2-\eta_{\text{min}})} \right)^2<0
\end{cases}
\end{equation} 
}

The minimum of $J^{'}(r)$ is reached if all but one $\|f_k\|$ norms are 0 if conditions \ref{eq:cond-fin} are simultaneously verified.
This happens for a well-chosen value of $\lambda{*}\in(0,\sqrt{\frac{3}{2}})$ and implicitly of $\beta_{*}\in(\eta_{\text{min}}+2;\eta_{\text{min}}+\frac{7}{2})$.

Remark that, as a function of $\beta_{*}$, $\left(2-\sqrt{\frac{3}{2}}\sqrt{\beta_{*}-2-\eta_\text{min}} \right)^2$ is decreasing when $\beta_{*}\in(\eta_\text{min}+2;\eta_\text{min}+\frac{7}{2})$. This means that we should chose the smallest value for $\beta_{*}$ which also verifies the first of conditions \eqref{eq:cond-fin}.

For an excess bandwidth $\gamma\in[0,0.1]$ a suitable value is $\beta_{*}=K(\beta_\text{min}-\frac{1}{2})+\frac{1}{2}$. Replacing this value in equations (\ref{eq:cond-fin}), we obtain conditions \eqref{ameliore:cond1'}.

\footnotesize{

}

\end{document}